\documentclass{aa}  
\usepackage{graphicx}
\usepackage{tabularx}
\usepackage{txfonts}
\usepackage{hyperref}

\hypersetup{colorlinks,linkcolor={blue},citecolor={blue},urlcolor={blue}}  
\usepackage{diagbox}
\usepackage{threeparttable}
\usepackage{lscape}
\usepackage{enumitem}
 
\usepackage{tikz,hyperref}
\usepackage{academicons}
\definecolor{lime}{HTML}{A6CE39}
\DeclareRobustCommand{\orcidicon}{%
	\begin{tikzpicture}
	\draw[lime, fill=lime] (0,0) 
	circle [radius=0.16] 
	node[white] {{\fontfamily{qag}\selectfont \tiny ID}};
	\draw[white, fill=white] (-0.0625,0.095) 
	circle [radius=0.007];
	\end{tikzpicture}
	\hspace{-2mm}
}

\foreach \x in {A, ..., Z}{%
	\expandafter\xdef\csname orcid\x\endcsname{\noexpand\href{https://orcid.org/\csname orcidauthor\x\endcsname}{\noexpand\orcidicon}}
}

\newcommand{\orcid}[1]{\href{https://orcid.org/#1}{\textcolor[HTML]{A6CE39}{\orcidicon}}}

\newcommand{\nicer}{\textit{NICER}\xspace}

\newcommand{\xmm}{{\it XMM-Newton}\xspace}

\newcommand{\swift}{{\it Swift}\xspace}

\newcommand{\nustar}{{\it NuSTAR}\xspace}

\newcommand{\ergs}[1]{$\times 10^{#1}$ erg s$^{-1}$}
\newcommand{\oergs}[1]{$10^{#1}$ erg s$^{-1}$}

\newcommand{\eqb}{\begin{eqnarray}}
\newcommand{\eqe}{\end{eqnarray}}

\graphicspath{{./}{Figs/}{Figs2/}}
\begin{document}

\title{Optical flares due to X-ray irradiation during BeXRB major outbursts} 
   \titlerunning{Optical flares during BeXRB major outbursts}
   \author{G. Vasilopoulos\orcid{0000-0003-3902-3915}\inst{1,2}\thanks{\email{gevas@phys.uoa.gr}}
          }
    \authorrunning{G. Vasilopoulos}
   \institute{
   Department of Physics, National and Kapodistrian University of Athens, University Campus Zografos, GR 15784, Athens, Greece
    \and
    Institute of Accelerating Systems \& Applications, University Campus Zografos, Athens, Greece
    }
   \date{Received ... ; accepted ...}

 
  \abstract
   {Be X-ray binaries (BeXRBs) may show strong X-ray and optical variability, and can exhibit some of the brightest outbursts that break through the Eddington limit. Major X-ray outbursts are often accompanied by strong optical flares that evolve parallel to the X-ray outburst.}
   {Our goal is to provide a simple quantitative explanation for the optical flares with an application to a sample of the brightest outbursts of BeXRBs in the Magellanic clouds and the Galactic Ultra-luminous X-ray (ULX) pulsar Swift J0243.6+6124.}
   {We constructed a numerical model to study X-ray irradiation in a BeXRB system. We then conducted a parametric investigation of the model parameters and made predictions for the intensity of the optical flares based on geometric and energetic constraints.}
   {From our modeling we found that the optical emission during major outbursts is consistent with being the result of X-ray irradiation of the Be disk. For individual systems, if this method is combined with independent constraints of the geometry of the Be disk, the binary orbital plane, and the plane of the observer, it can provide estimates of the Be disk size during major outbursts. Moreover, we computed a semi-analytical relation between optical flare luminosity and X-ray luminosity that is consistent with both model predictions and observed properties of flares.} 
   {}    
   \keywords{Stars: emission-line, Be -- Stars: neutron -- (Stars:) pulsars: general -- (Galaxies:) Magellanic Clouds -- X-rays: binaries}
   \maketitle
%

\section{Introduction}

High-mass X-ray binaries (HMXBs) are young binary systems composed of a compact object and a massive star, and they are commonly bright in X-rays due to the large mass transfer rates.
Many HMXBs have a Be star donor and are referred to as BeXRBs \citep[for a review see][]{Reig2011}. BeXRBs that host neutron stars (NSs) are some of the most variable systems in the X-ray sky.
Be stars are characterized by infrared excess and strong emission lines, which in term are highly variable in intensity and shape \citep[][]{Harmanec1982}. It is generally accepted that these lines originate from a ``decretion" disk of ejected ionized gas. Long-term variability in the optical implies a constantly evolving disk. At the same time, strong X-ray emission from BeXRBs occurs during type I and major type II outbursts. Type I X-ray outbursts ($\mathrm{L_{X} \sim 10^{36}-10^{37} erg \: s^{-1}}$) are typically periodic and occur near the periastron passage, while  Type II outbursts ($\mathrm{L_{X} \gtrsim 10^{37} erg \: s^{-1}}$) last multiple orbital periods and can occur every tens of years or more.

BeXRBs are also variable in the optical wavelengths. Periodic modulation in time scales of days may be related to non-radial pulsations of the Be star. Periodicity in timescales of weeks to months is often proven to be related to orbital effects, while there are a couple high eccentric systems with periods of many years \citep[e.g.][]{2017A&A...598A..69H,2018MNRAS.474L..22P}.
However, a general feature of Be stars is long-term variability on the order of thousands of days \citep[e.g.,][]{Kourniotis2014,Treiber2021,2025A&A...694A..43T}, which can be due to disk depletion and formation events \citep[e.g.,][]{deWit2006} and also be associated to disk precession and truncation effects driven by the NS \citep{Martin2011,Martin2023}.

In the era of multi-wavelength astronomy, most major outbursts of BeXRBs have been monitored in X-ray and optical. In particular, in the Magellanic Clouds (MCs), a handful of BeXRBs had Type-II X-ray outbursts with $L_{\rm X}$ larger than \oergs{38} within the last 10 years. Another great example is the major 2017 outburst of Swift J0243.6+6124 which is labeled as one of the closest Ultra-luminous X-ray sources \citep{2018ApJ...863....9W}. All these bright X-ray events are associated with strong optical flares that may not directly be related to the long-term super-orbital variability commonly seen in Be stars. 

Optical variability and flares has been the focus of numerous studies in X-ray binaries revealing all short of intriguing phenomena. Reflection and irradiation has been studied in black hole and neutron star binaries \citep{2001MNRAS.320..177W}, from ms \citep{2010MNRAS.407.2166G,2023Natur.615...45V} to long timescales \citep{2001MNRAS.320..177W,2005MNRAS.362...79C}. 
Such studies have recently been extended to multi-wavelength properties of Ultra-luminous X-ray sources \citep[e.g.][]{2022MNRAS.509.4694A}.
The inclusion of a geometrical structure around Be stars, offers another complexity to such modeling which is worth further investigating.  

In this work, we have identified a sample of such optical flaring events in the recent literature associated with major Type-II outbursts \citep[e.g.][]{2025A&A...694A..43T}. To explain these optical flares quantitatively, we have constructed a numerical model that computes the irradiation effects on the Be star and the disk based on a given geometrical configuration. 
We will present the model predictions for individual systems and how this can be combined with other independent measurements of the orbital parameters to constrain the Be disk properties. 
Finally, we will discuss how this model predicts a linear scaling between the logarithms of the luminosities of the optical flares and X-ray luminosity and how this compares with observational data.
\section{Data and Sample selection}

\begin{table*}
\caption{BeXRBs observed parameters.}\label{tab:bexrbs}
\centering
\begin{tabular}{lccccccc}
\hline
\hline
\multicolumn{8}{c}{BeXRBs - literature values}\\
\hline
Name & Distance & Date$^{(a)}$ & $L_{\rm X}$ $^{(b)}$ &  P$_{\rm orb}$ / e / $a\sin{i}$ $^{(c)}$ & Spectral Class & A$_I$$^{(d)}$ &Refs.$^{(e)}$ \\
 & (kpc) & (MJD) & ($10^{38}$ erg/s) & (d) / -- / (ls) & & & \\
\hline
LXP\,8.04 & 50 & 56675 & 4  & 23.97 / 0.037 / 105  & O9Ve &  0.0885& [I]\\ 
SMC\,X-2 & 60  & 57300 & 7 & 18.38 / 0.07 / 73.7&   O9.5 III–V &  0.0735& [II] \\ 
SMC\,X-3 & 60  & 57630 & 25 & 45.07 / 0.231 / 189  & B1–B1.5IV–V&  0.066& [III]\\ 
SXP\,4.78 & 60  & 58450 & 1.8* & 20-100 / -- / -- & B1–3IIIe& 0.144&[IV]\\ 
SXP\,5.05 & 60  & 56556 & 1.3* & 17.13 / 0.155 / 142.4 & B0.2Ve& 0.114& [V]\\ 
Swift\,J0243.6+6124 & 5.5  & 58067 & 20 & 27.7 / 0.103 / 115.5 & O9.5Ve& 1.9& [VI]\\
\hline
\end{tabular}
\tablefoot{
\tablefoottext{a}{Date in Modified Julian Date (MJD).}
\tablefoottext{b}{Maximum bolometric X-ray luminosity in units of $10^{38}$ erg/s. Values marked with * are obtained from soft band using standard corrections (see text).}
\tablefoottext{c}{Orbital parameters:  Period in d, eccentricity, projected semimajor axis in light-seconds.}
\tablefoottext{d}{Extinction in I band from for the MC sources \citep{2021ApJS..252...23S}, and J0520 \citep{2020A&A...640A..35R}.}
\tablefoottext{e}{Parameter references: 
[I] \citet{2001MNRAS.324..623C,2014A&A...567A.129V,2023MNRAS.520..281K}, [II] \citet{2008MNRAS.388.1198M,2011MNRAS.416.1556T,2023MNRAS.521.3951J}, [III] \citet{2017A&A...605A..39T}, 
[IV] \citet{2018ATel12219....1G,2018ATel12224....1M,2019MNRAS.485.4617M}, 
[V] \citet{2015MNRAS.447.2387C}, 
[VI]  \citet{2018ApJ...863....9W,2023MNRAS.520..281K} }
}
\end{table*}

\subsection{Optical data from BeXRB outbursts}\label{ogle}

Historically, multiple outbursts of BeXRBs have been observed in the optical by the Optical Gravitational Lensing Experiment \citep[OGLE,][]{1992AcA....42..253U, 2008AcA....58...69U,Udalski2015}, which has monitored the MCs since 1992. 
Since the start of OGLE-II in 1997, images have been taken at Las Campanas Observatory in V- and I-bands with the 1.3\,m Warsaw telescope. The subsequent phases OGLE-III and OGLE-IV involved improved CCDs and an increased field of view. The OGLE database\footnote{OGLE XROM portal: \url{https://ogle.astrouw.edu.pl/ogle4/xrom/xrom.html}} provides the light curves of most known X-ray binaries in the MCs and \citet{2008AcA....58...69U} describes the data reduction. 

\subsection{X-ray data from outbursts}

Major outbursts in the MCs are often detected by all-sky surveys from observatories like BAT on the Neil Gehrels Swift Observatory \citep[i.e. \swift][]{2004ApJ...611.1005G}. Follow-up observations are typically performed by \swift/XRT, \nicer or \xmm in soft X-rays (0.3-10 keV). The bolometric $L_{\rm X}$ is best estimated from broadband X-ray data obtained by telescopes like \nustar \citep{2013ApJ...770..103H}, otherwise it may be inferred using empirical scaling relations. For Type-II outbursts $L_{\rm X}$ in 0.5-10 keV band is about 40\% of the bolometric $L_{\rm X}$ \citep{2022MNRAS.513.1400A}. X-ray analysis of the data is beyond the scope of this work, and we will rely on data from the literature for $L_{\rm X}$ estimates.

\subsection{BeXRB major outbursts}

\begin{figure*}[h]
    \centering
    \includegraphics[width=\columnwidth]{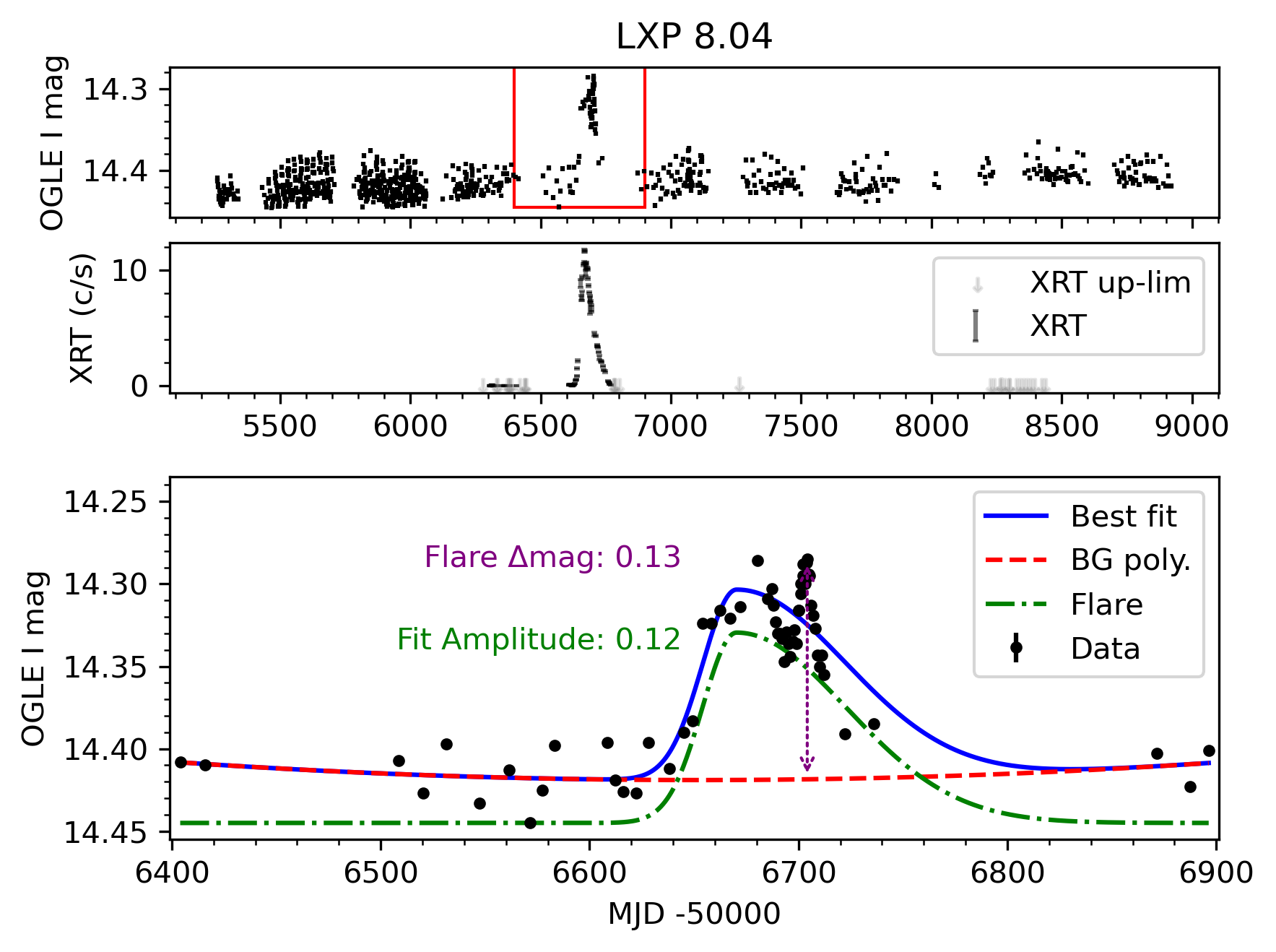}
    \includegraphics[width=\columnwidth]{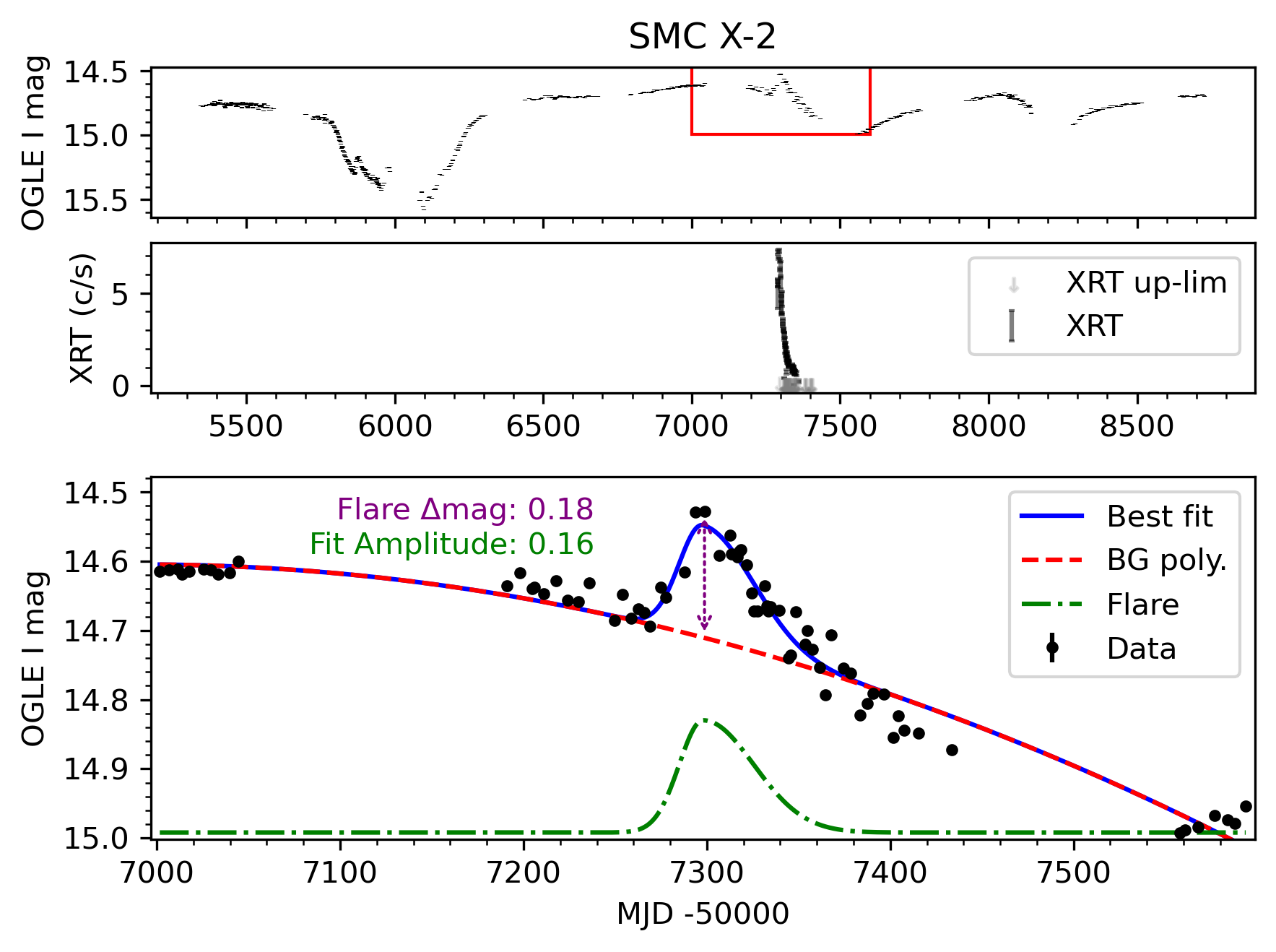}\\
     \includegraphics[width=\columnwidth]{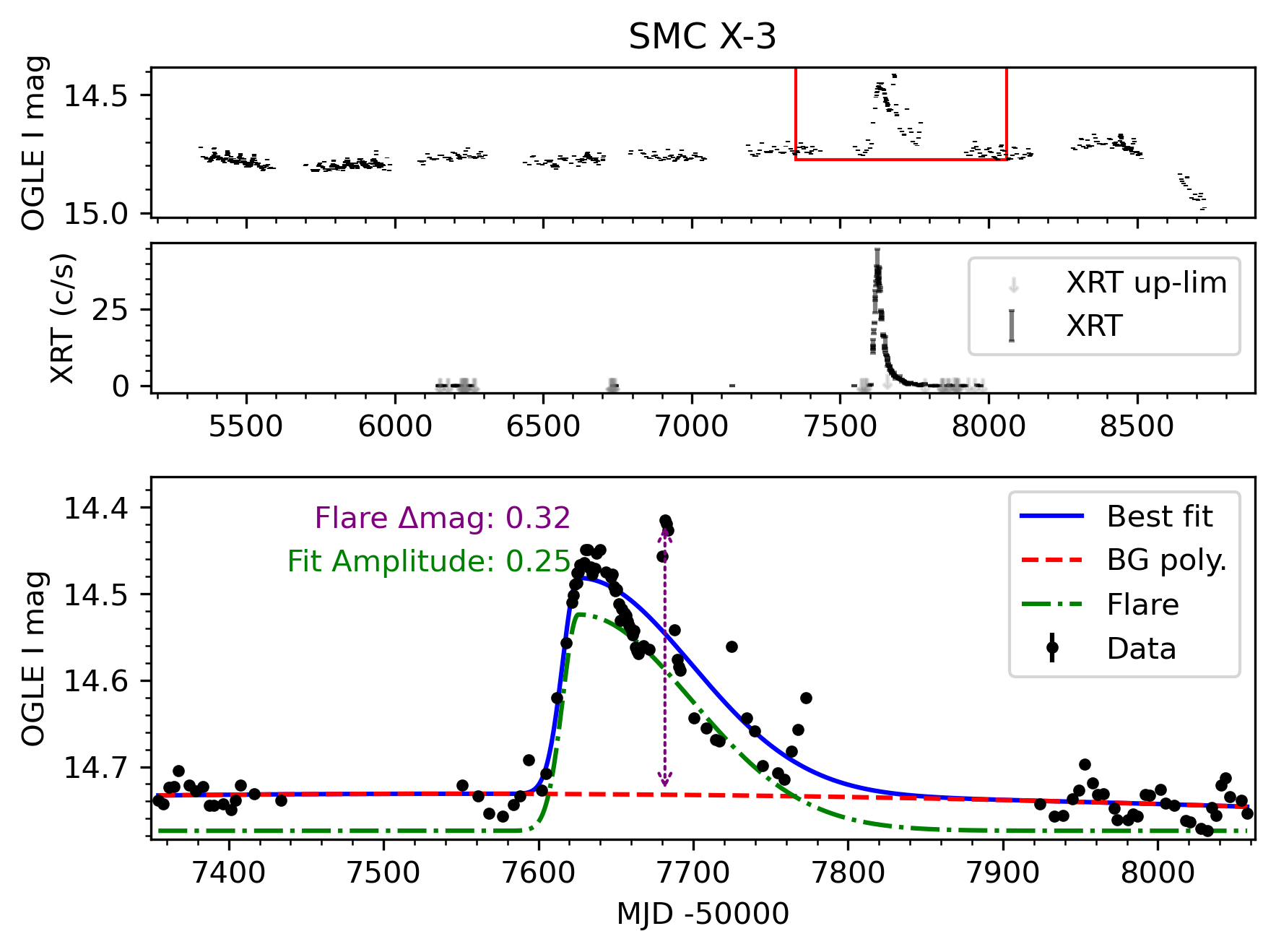}
      \includegraphics[width=\columnwidth]{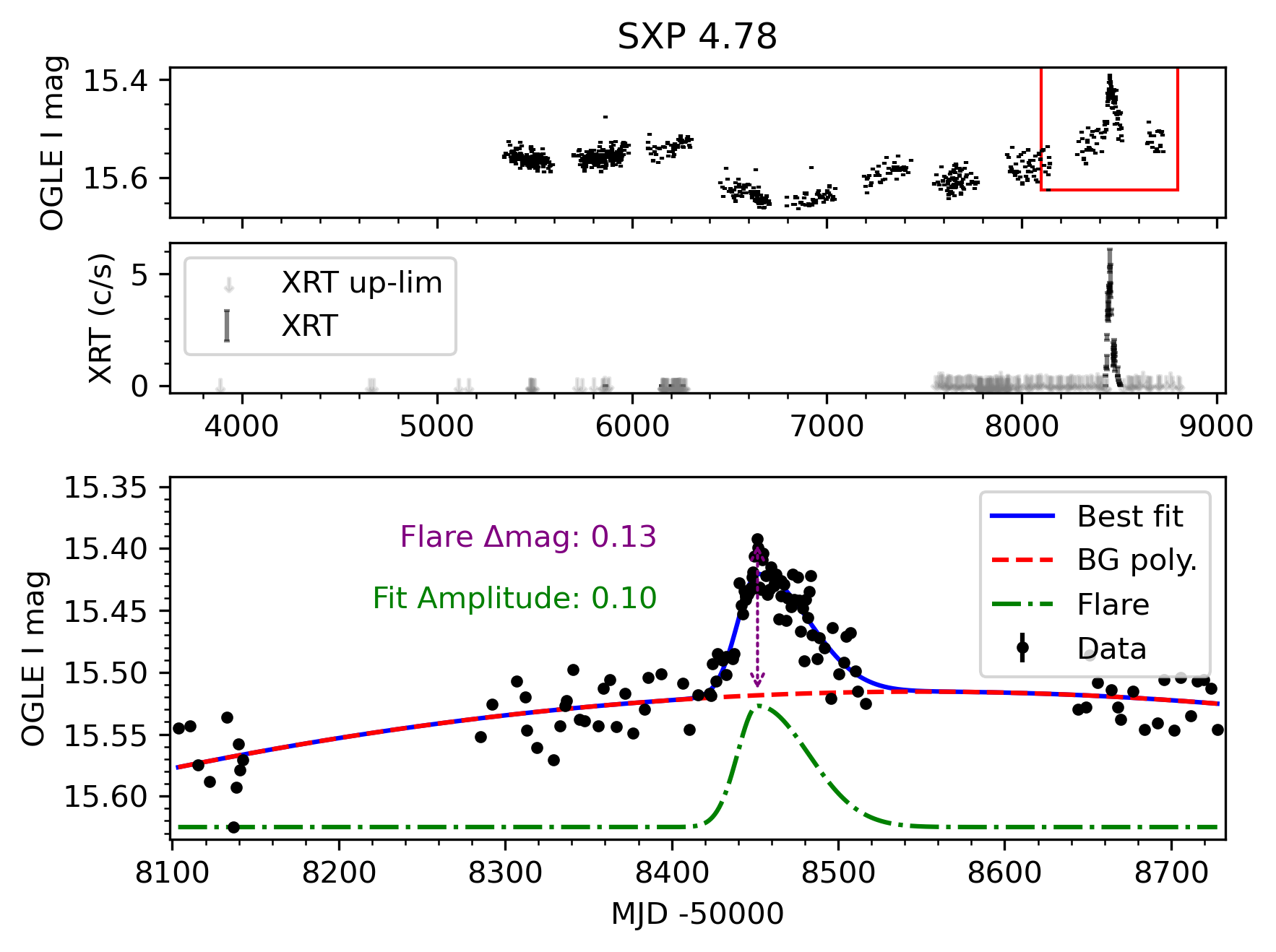}\\
      \includegraphics[width=\columnwidth]{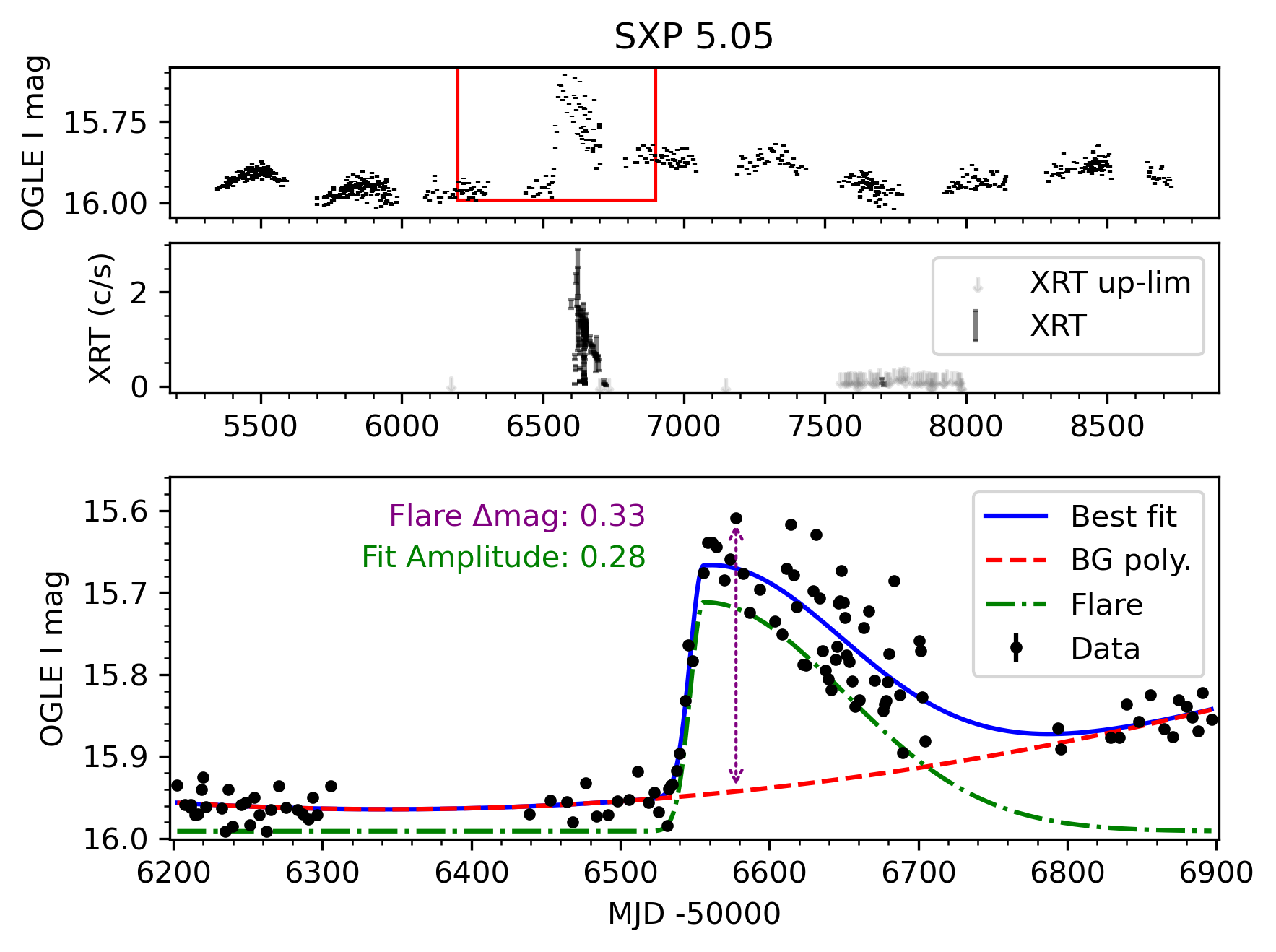}
      \includegraphics[width=\columnwidth]{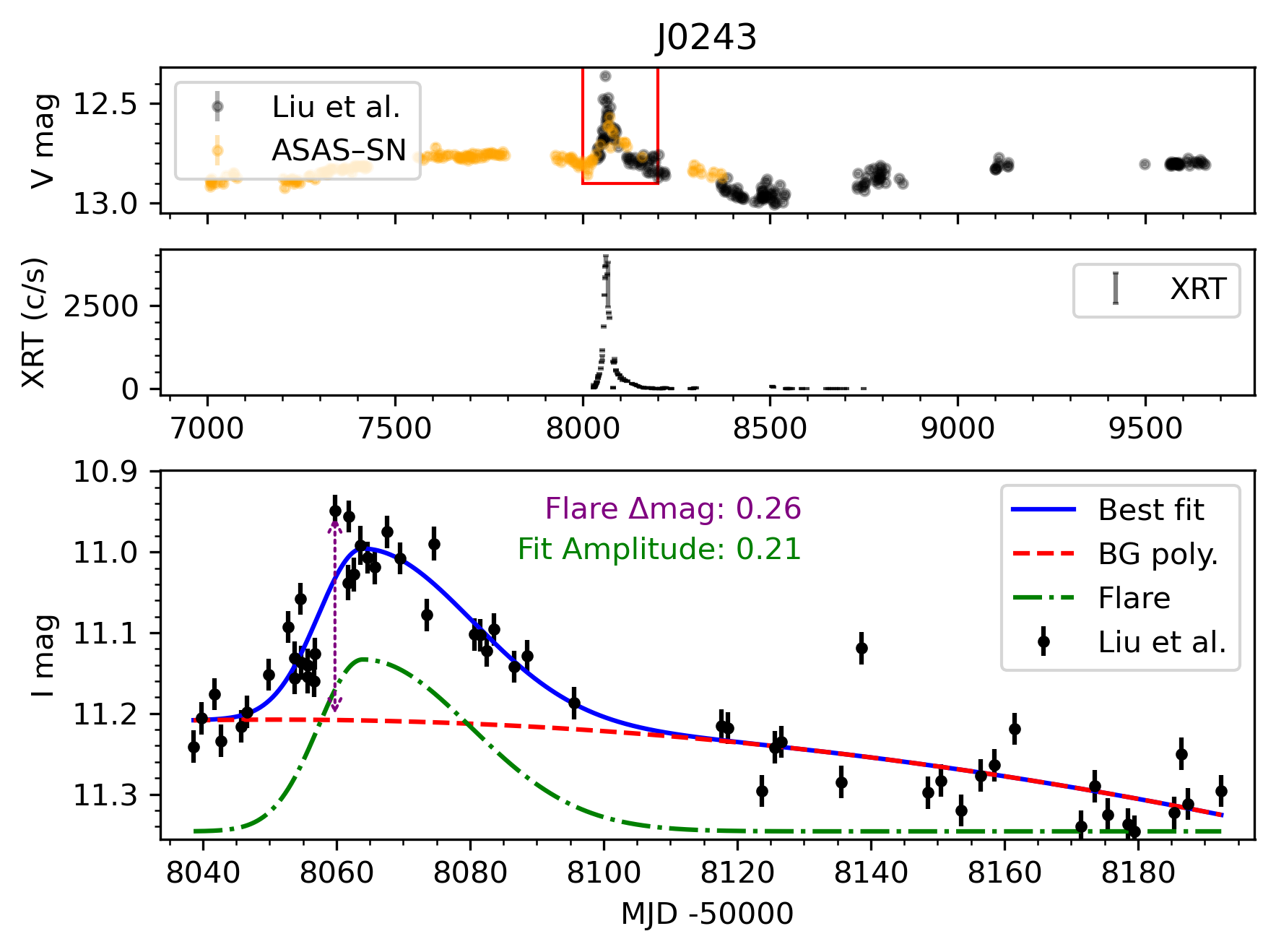}
      \vspace{-0.2cm}
    \caption{Optical flares during major outbursts from BeXRBs. The top panels show the historic OGLE I band light curves, while the epoch of flares are marked with red boxs. Middle panels present \swift/XRT 0.3-10.0 keV light-curves. Each flare (see lower panel) is fitted by a model composed of a polynomial and a two-sided Gaussian. The flare amplitude from the fit is given in the label with green fonts. We also mark the I mag of maximum flux with a magenta vertical arrow and provide the $\Delta{mag}$ difference of that point from the polynomial background. The difference between the two maxima could be related to orbital modulation during the flares (see the cases of SXP\,5.05 and LXP\,8.04).
    }
    \label{fig:flares}
\end{figure*}

For our study, we will focus on 6 BeXRB outbursts, from 4 systems located in the SMC, one in the LMC, and one Galactic system as presented in Table \ref{tab:bexrbs}. Regarding our selection criteria, we searched for bright outbursts in the MCs having peak luminosity higher than \oergs{38} that also had available optical monitoring data via OGLE (see sec. \ref{ogle}). For example one of the brightest outbursts in the LMC was that of RX J0209.6-7427 \citep{2020MNRAS.494.5350V}, but due to its location, there was no OGLE coverage. The selected systems are:
\begin{itemize}[left=0pt]
    \item RX\,J0520.5-6932: also known as LXP\,8.04 is an X-ray pulsar in the LMC \citep{2014A&A...567A.129V}. Its 2014 Type-II outburst was monitored with \swift/XRT, while 2 \nustar visits were performed \citep{2014ApJ...795..154T}.
    \item SMC\,X-2: Its 2015 outburst was monitored by \swift/XRT while 3 \nustar pointings were performed \citep[e.g.]{2023MNRAS.521.3951J}.
    \item SMC\,X-3: Its 2016-7 major outburst was perhaps the brightest one from a pulsar in the SMC \citep{2017A&A...605A..39T}. It was monitored by \swift/XRT while 2 \nustar pointings were performed during the outburst.
    \item SXP\,4.78: The 2018 outburst of the system was monitored by \swift/XRT and \nicer showing a peak luminosity of about 7\ergs{37} at 0.5-8 keV \citep{2018ATel12219....1G}, which would translate to a bolometric value of 1.8\ergs{38} assuming typical correction values. No orbital period is known for the system, however empirical relations put the period between 20-100 d. A single peak H$_{\alpha}$ emission line would indicate a face-on Be disk orientation \citep{2019MNRAS.485.4617M}.
    \item SXP\,5.05: \citet{2015MNRAS.447.2387C} have studied extensively this intriguing system. It shows X-ray eclipses from the Be disk, indicating an almost edge-on orbital modulation. A double peak H$_{\alpha}$ emission line also indicates an edge-on disk. 
    The X-ray luminosity of the system in the 0.5-10 keV was 0.5\ergs{37} indicating a peak bolometric $L_{\rm X}$ of 1.3\ergs{38}.
    \item Swift J0243.6+6124: This source (hereafter J0243) is refered to as the Galactic ULX as it reached \oergs{39} during its 2017 major outburst \citep{2018ApJ...863....9W}. Regarding its distance many early works in the literature adopted a value of 7 kpc while GAIA DR3 provide the updated value of 5.2 kpc \citep{2021AJ....161..147B}. The detailed orbital solution has been obtained for the system \citep[e.g.][]{2023MNRAS.520..281K}. The optical companion is a O9.5Ve star indicating a 30$^o$ inclination of the orbital plane compared to the observer \citep{2020A&A...640A..35R}. For the optical evolution of the outburst, we used published \citep{2022A&A...666A.110L} and public data \footnote{ASAS-SN: \url{https://asas-sn.osu.edu/photometry/71ac277e-d7a0-5387-9385-7be7ba02931f}} in the V band. 
For the overall evolution of J0243 we looked at the V band light-curve that includes data from \citet{2022A&A...666A.110L} and ASAS-SN. Due to the presence of a contaminating source in the field we introduced a 0.2 mag shift in ASAS-SN magnitudes \citep[see][for details about the optical data]{2022A&A...666A.110L}. However, for consistency in modeling the flare, we used only I-band light curves. 
\end{itemize}

For our sample all optical light curves are presented in Fig. \ref{fig:flares}, while for comparison we also plot X-ray data from 
\swift/XRT to pinpoint the epoch of the X-ray flares. In regards to optical magnitudes, it is important to correct for extinction. For the MC systems corrections are minimal and can be performed by using literature values for reddening, where $A_I\approx1.5E(V-I)$ \citep{2021ApJS..252...23S}. For J0243 we adopted an $A_V=3.48$ value \citep[][]{2020A&A...640A..35R} and extinction curve of \citet{2007ApJ...663..320F} to compute $A_I$. The values are presented in Table \ref{tab:bexrbs}.

\subsection{Modeling of optical light curves}

Our main goal is to model the optical light curves and measure the intensity of the optical flares. 
After identifying the optical flares we perform a fit that combines a quadratic polynomial background and a two-sided Gaussian function. The flare function is expressed as:
\begin{equation}
\text{Flare}(t) = - A \exp\left(-\frac{(t - t_{\rm peak})^2}{2\sigma_{1,2}^2}\right), \quad
\begin{cases}
\sigma_1, & t < t_{\rm peak} \\
\sigma_2, & t \ge t_{\rm peak}
\end{cases}
\end{equation}
where $t_{\rm peak}$ is the peak time of the flare, $A$ its amplitude in magnitude units and $\sigma_{1,2}$ the rise and decay term of the flare.
The above method can be used effectively for smooth flares, however, in many cases there are sharper features in the light-curve that last only a few days. To study the effect of these mini-flares we also identified the maximum of the flare and compared it with the background level estimated from the fit. Then the maximum of these mini flares is $\Delta m = m_{\text{bg}} - m_{\text{flare}}$. Since these flares are compared to the background the $\Delta m$ value is actually larger than $A$ value. We note that these mini-flares are not crucial to the analysis perform from this point on, but are useful to identify the effect of noise to the data from this secondary variability on top of the major optical flares.
The results of this modeling are presented in Fig. \ref{fig:flares} and Table \ref{tab:flares}. For systems like  SXP\,5.05 and LXP\,8.04 there is still strong orbital modulation during the flaring period, which affects the flare shape, thus in these cases the $\sigma_{1,2}$ values derived from the fit are unreliable. In any case, although we list all parameters from the fit, our emphasis is only on the amplitude of the flare and all other parameters should be used with caution. 

\begin{table}
\caption{BeXRBs optical flare parameters.}\label{tab:flares}
\centering
\begin{tabularx}{\columnwidth}{lccccc}
\hline
\hline
\multicolumn{6}{c}{BeXRBs Flares}\\
\hline
Name & I$_{BG}$$^{(a)}$   &$\Delta \rm mag$ & A [mag]& $\sigma_1$ [d]& $\sigma_2$ [d]\\
\hline
LXP\,8.04 & 14.42 & 0.13 & 0.112$\pm$0.010& 13$\pm$5 & 57$\pm$12\\ 
SMC\,X-2 & 14.71 & 0.18 & 0.16$\pm$0.01& 13$\pm$3 & 28$\pm$4\\
SMC\,X-3 & 14.73 & 0.32& 0.25$\pm$0.01& 10$\pm$3 &  73$\pm$6\\
SXP\,4.78 & 15.52 & 0.13 & 0.098$\pm$0.005&  12$\pm$2&  30$\pm$3\\
SXP\,5.05 &  15.95 & 0.33& 0.279$\pm$0.014&  9$\pm$3& 94$\pm$6\\
J0243 & 11.2 &0.26 & 0.213$\pm$0.020& 7$\pm$2& 16$\pm$3\\ 
\hline
\end{tabularx}
\tablefoot{
\tablefoottext{a}{Estimated from BG level at peak of flare.}
}
\end{table}

\section{The irradiation model}

\subsection{Setup of the X-ray irradiation model}

We have constructed a numerical model to compute the temperature change, the change in emitted flux, and the optical magnitude of the Be Star and its surrounding decretion disk due to the X-ray irradiation from an X-ray source (i.e. the accreting NS). For the geometrical problem, we use a Be star, a Be disk, and the X-ray source placed at a distance $R_{sep}$. 
We create a grid of points to map the surface of a spherical star with radius $R_{\rm Be}$ as well that of a flat disk that we treat as a 2-dimensional surface with inner and outer radius $R_{in}$ and $R_{out}$. The X-ray luminosity $L_{\rm X}$ and the inclination $\theta_{NS}$ of the NS compared to the plane of the Be disk is also a free parameter. A $\theta_{NS}$ value of zero denotes an edge-on disk compared to the NS position, while 90 deg a face on disk.
For the non X-ray irradiated system we can adopt a star temperature $T_\star$ and a 
disk temperature profile that drops as a function of distance, following a typical relation:
\begin{equation}
T_{\rm disk}(r) = T_{\rm disk, 0} \left( \frac{r}{R_{\star}} \right)^{-n},
\label{eq2}
\end{equation}
where $T_{\rm disk, 0}$ is the disk temperature at $R_{\star}$ and $n$ takes typical values of $-3/4$ \citep[e.g.][]{2006ApJ...639.1081C}. 
Observational applications to BeXRBs disk variability follow the above temperature profile with $T_{\rm disk, 0}=0.9T_\star$ \citep[e.g.][]{deWit2006}.
However, close to the star detailed integration of non-LTE codes may yield a more complex behavior \citep{2006ApJ...639.1081C}: 
\begin{equation}
T_{\rm disk}(r)  = 
\frac{T_\star}{\pi^{1/4}}
\left[
\sin^{-1}\left( \frac{R_\star}{r} \right) - 
\frac{R_\star}{r}
\sqrt{1-\frac{R_\star^2}{r^2}}
\right]
^{1/4}.
\label{eq3}
\end{equation}
The above equation provides a good agreement with simulations but can deviate for large $r$ values or very small disk densities \citep[see][for discussion]{2006ApJ...639.1081C}.
For our problem and typical disk configuration (i.e., \( r \in [1.5, 7] R_\star \)), eq. \ref{eq2} and eq. \ref{eq3} agree for $n\approx0.8$ and $T_{\rm disk, 0}\approx0.72T_\star$. For the rest of the paper we will adopt eq. \ref{eq3}.

For a particular orientation, we compute the distance of each surface element of the grid (for the Be star and the disk) from the X-ray source $( d_{\rm ele})$, and the cosine of the angle between the normal of the surface element and the direction of the NS, $\cos \phi_{\rm ele}$. Then the temperature change $\Delta T$ is computed from the following expression:
\begin{equation}
T_{\rm new}^4 = T_{\text{initial}}^4 + \frac{(1 - a) L_{\rm X} \cos \phi_{\rm ele.}}{4 \pi \sigma d_{\rm ele}^2},
\end{equation}
where $a$ is the albedo of the star or disk which we assume to be 0.5 and $\sigma$ is the Stefan-Boltzmann constant.

\begin{figure}
    \centering
    \includegraphics[width=\columnwidth]{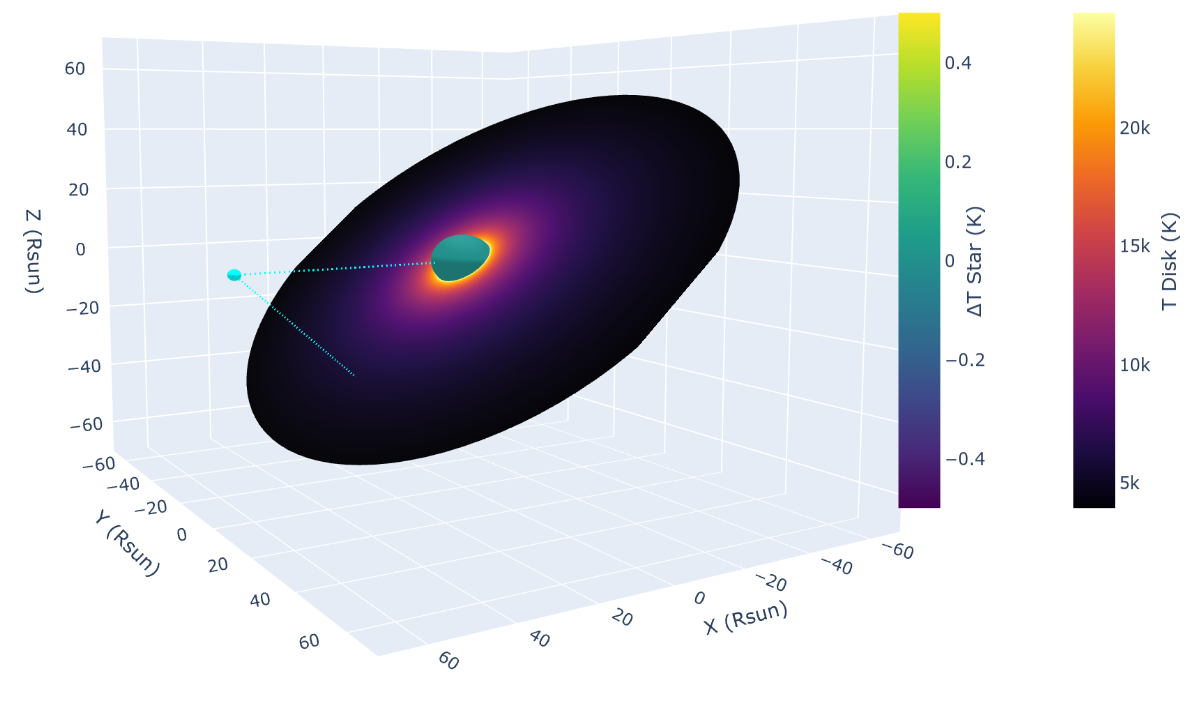}
     \includegraphics[width=\columnwidth]{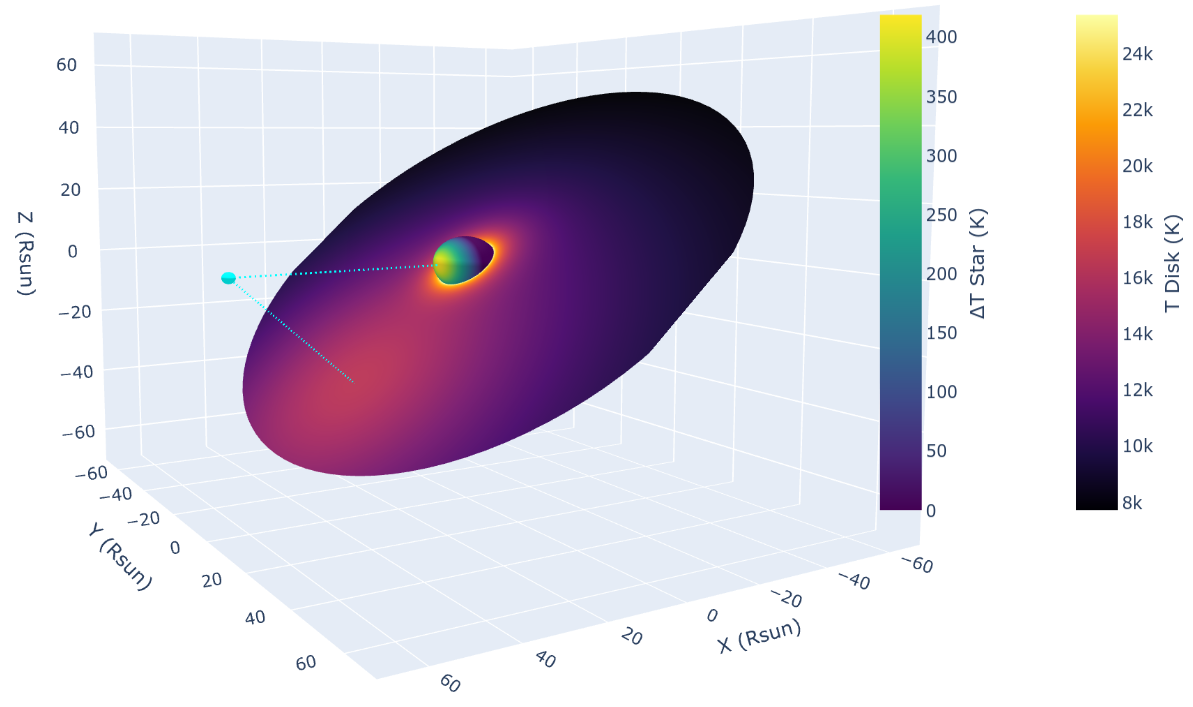}
      \includegraphics[width=\columnwidth]{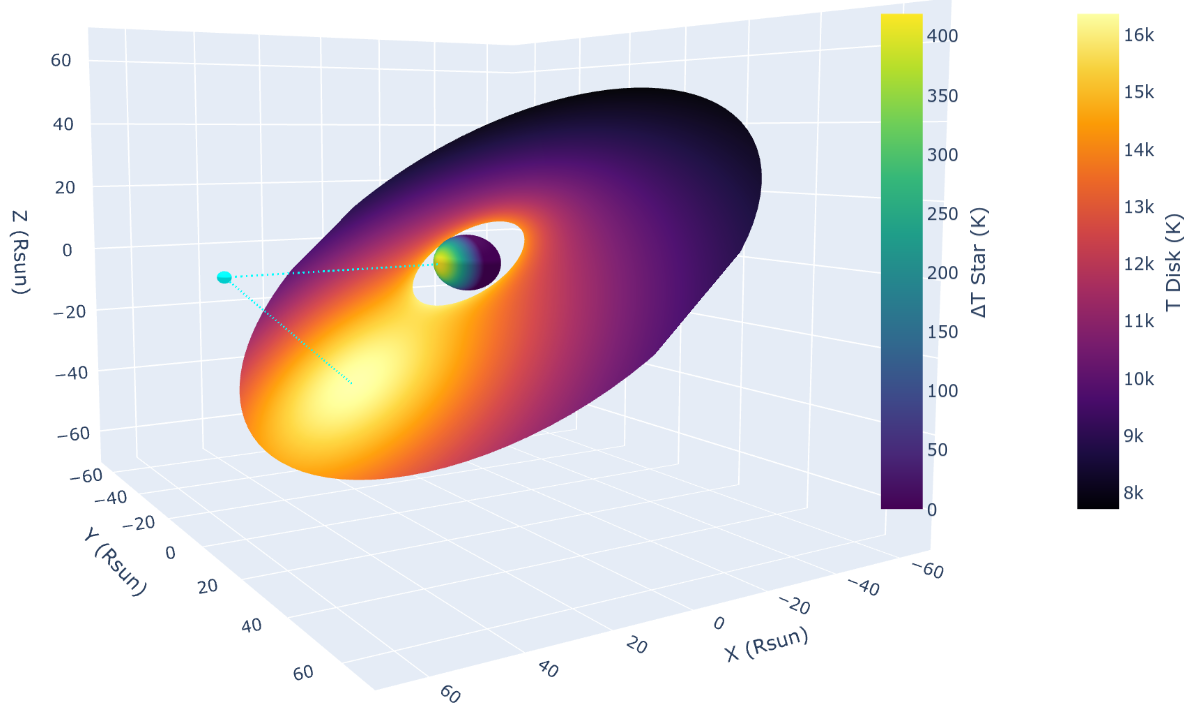}
    \caption{Examples of model setup, all objects are to scale apart from the NS size (cyan color). From top to bottom we used 3 different sets For $L_{\rm X}$ and $R_{\rm in}$; $[0, R_{\rm Be}]$,  $[10^{39} erg/s, R_{\rm Be}]$ and $[10^{39} erg/s, 2R_{\rm Be}]$. For all cases $T_0$ was set at 30,000 K, $R_{\rm out}$ was set at $9R_{\rm Be}$, and inclination at 45$^o$. Colors indicate $\Delta T$ compared to the minimum temperature of each object element and are computed separately for disk and star for illustration purposes. If the Be disk touches the Be star surface, then its maximum temperature is at that point, regardless of irradiation or not. 
    }
    \label{fig:Bedisks}
\end{figure}

In Fig. \ref{fig:Bedisks} we plot 3 different problem setups to illustrate the geometry of the problem and the effect of irradiation. In this setup, we place the NS at a distance of $R_{sep}\sim64R_{\odot}$ and $R_{\rm Be}=8R_{\odot}$, to roughly represent systems like LXP\,8.04 and J0243. 
For all cases $T_0$ was set at 30,000 K, $R_{\rm out}$ was set at $9R_{\rm Be}$, and inclination angle $\theta_{NS}$ at 45$^o$.
From these plots, we may extract a few quantitative takeaways. The surface of the Be star is not significantly heated by irradiation, and even for \oergs{39} the surface temperature only changes by $\sim$1-2\%. Moreover, the effect of heating in the disk should be more evident, in terms of change of magnitude, for systems where the inner radius of the Be disk is truncated away from the Be star surface. When the inner Be disk radius touches the Be star then even with irradiation, the maximum temperature of the disk is at $R_{\rm Be}$, while when the inner Be disk radius is $2R_{\rm Be}$, the maximum temperature after irradiation is at the point closest to the NS. We note that such large Be disks as in our examples would not be dynamically stable as they would be truncated by dynamical forces.

To compute the spectral energy distribution (SED) from the irradiated disk, we sum the contributions from each differential area element \( \Delta A \) on the disk. For a face-on view, the total luminosity \( L_\lambda \) at a given wavelength \( \lambda \) is obtained by summing the specific intensities over all the elements:
\begin{equation}
L_\lambda = \sum_{i} \pi B_\lambda(\lambda, T_{i}) \Delta A_{i},
\end{equation}
where \( B_\lambda(\lambda, T_{i}) \) is the Planck function at temperature \( T_{i} \) corresponding to the \( i \)-th element, and \( \Delta A_{i} \) is the area of the \( (i) \)-th element. The factor of \( \pi \) accounts for the integration over all solid angles for isotropic emission. This formulation provides the SED as observed from a viewpoint directly above the disk, assuming a face-on orientation for the observer. For the spectral energy distribution (SED) measured by an observer at a distance $D$ we compute the flux by dividing by $4\pi D^2$ and $2\pi D^2$ for the star and disk respectively, assuming that the disk emits only from the irradiated side.

\subsection{Optical magnitudes}

\begin{figure}
    \centering
    \includegraphics[width=\columnwidth]{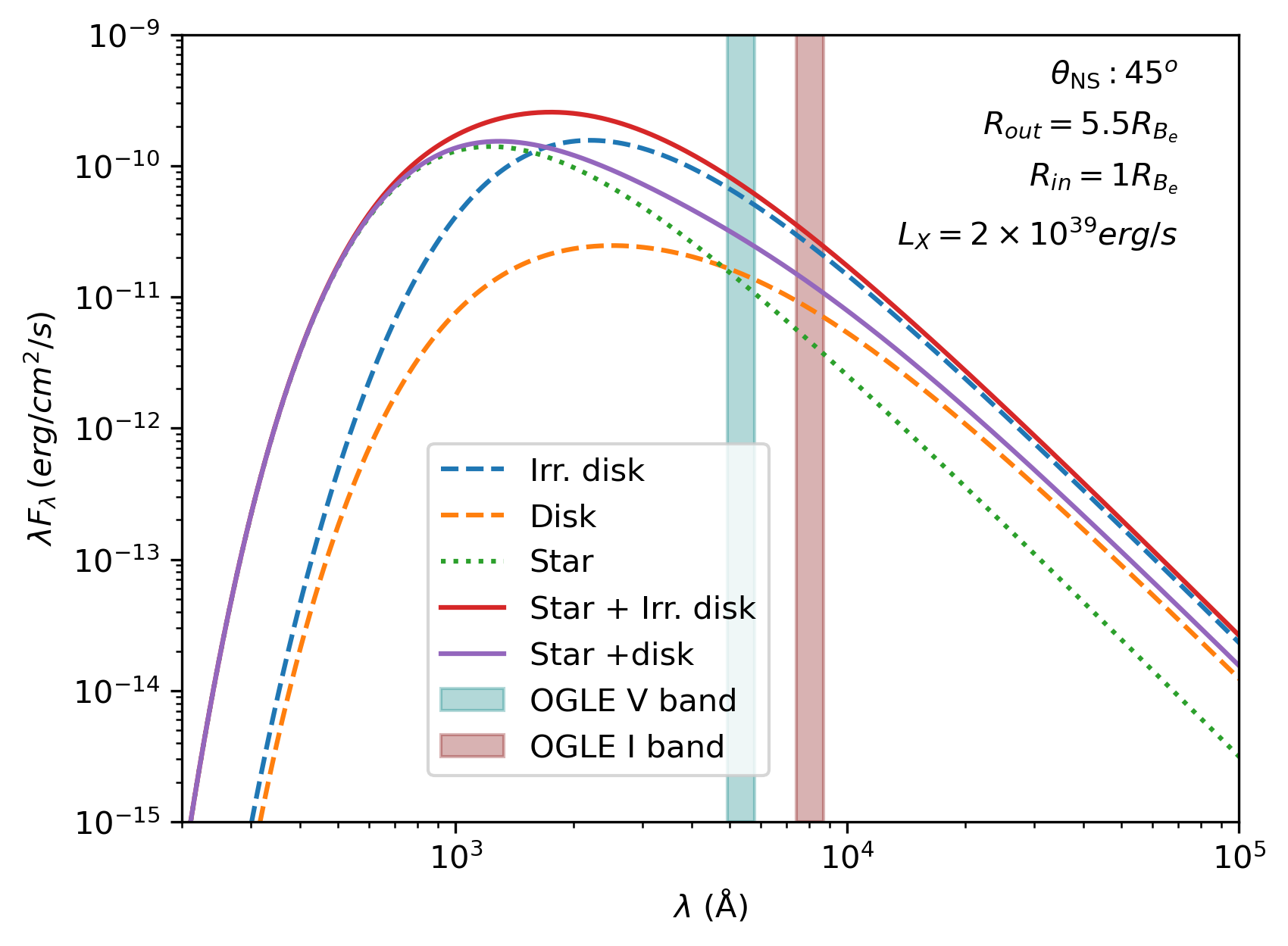}
    \caption{Computed SED for a particular system configuration, similar to the ones presented in Fig. \ref{fig:Bedisks}. Note the lower value of $R_{\rm out}$ for the Be disk to encapsulate a truncated outer disk. We plot the total SED and individual components 
    of the star, the non-irradiated disk, and the irradiated disk. Vertical stripes mark the range of the OGLE filters.}
    \label{fig:sed}
\end{figure}

After computing a simulated SED (i.e. \( F(\lambda) \)) we need to estimate the fluxes and magnitudes in the optical filter. We will focus in OGLE optical filters that include V and I pass-bands. First we take into account the filter transmission curves (e.g. \( T_V(\lambda) \) and \( T_I(\lambda) \)) and camera quantum efficiency (\( QE(\lambda) \))\footnote{Available through the OGLE project \url{https://ogle.astrouw.edu.pl/main/OGLEIV/mosaic.html}}: 
\begin{equation}
F_{V,I} = \frac{\int_{\lambda_{\text{min}}}^{\lambda_{\text{max}}} QE(\lambda) \cdot T_{V,I}(\lambda) \cdot F(\lambda) \cdot \lambda \, d\lambda}{\int_{\lambda_{\text{min}}}^{\lambda_{\text{max}}} QE(\lambda) \cdot T_{V,I}(\lambda) \cdot \lambda \, d\lambda},
\end{equation}
and then we compute the magnitudes in the standard Johnson-Cousins systems\footnote{Note that for zero mag calibration corrections $f_\lambda$ and $f_\nu$ are reversed in A2 table of \citet{1998A&A...333..231B}} \citep{1998A&A...333..231B}.  An example of a theoretical SED from the Be star and disk system, and its decomposition in various spectral components is presented in Fig. \ref{fig:sed}. We note that irradiation of the Be star would only increase the temperature of $\sim1\%$ and a total increase in luminosity of less than $2\%$ given that only part of its surface is heated, thus it has no effect on the SED even at high X-ray luminosity level. A final note should be on the numerical accuracy of our method. We used a grid of $N\times{N}$ points for the disk and star surface and experimented with grid sizes. We found that this mainly affected the peak temperature of each surface and affected the SED mostly on higher energies. For example, using $N=100$ vs $N=300$ for our grid only introduced a 1-3$\%$ difference in computed luminosity in the I energy band. We thus adopted a value of $N=100$ throughout this work.

\section{Results}

\subsection{Application to individual BeXRBs}

We will now perform simulations for individual BeXRB outbursts taking into account the orbital separation and observed $L_{\rm X}$.
Regarding the optical companions of our sample, all have spectral types consistent with B1-O9, and luminosity class V apart from SXP\,4.78. 
Thus their temperatures are between 26000-34000 K \citep{2013ApJS..208....9P}, for simplicity for our calculations we adopt 30000 K for all systems. The Be star should have a radius of $8$R$_{\odot}$ similar to J0243 \citep{2020A&A...640A..35R}.
Here we will present an application to SXP\,5.05 while our results for the other systems are presented in Appendix \ref{sec:app}.
\citet{2015MNRAS.447.2387C} measured the binary orbital parameters and put constraints on the Be star radius ($\sim5.2\times10^{11}$ cm), binary separation ($\sim4.08\times10^{12}$ cm), and Be disk size ($2.6\pm1.8\times10^{12}$ cm). Based on the eclipses they argued for a $\theta_{\rm LOS}\sim83^\circ$ observer line of sight compared to the normal of the orbital plane.

\begin{figure}
    \centering
    \includegraphics[width=\columnwidth]{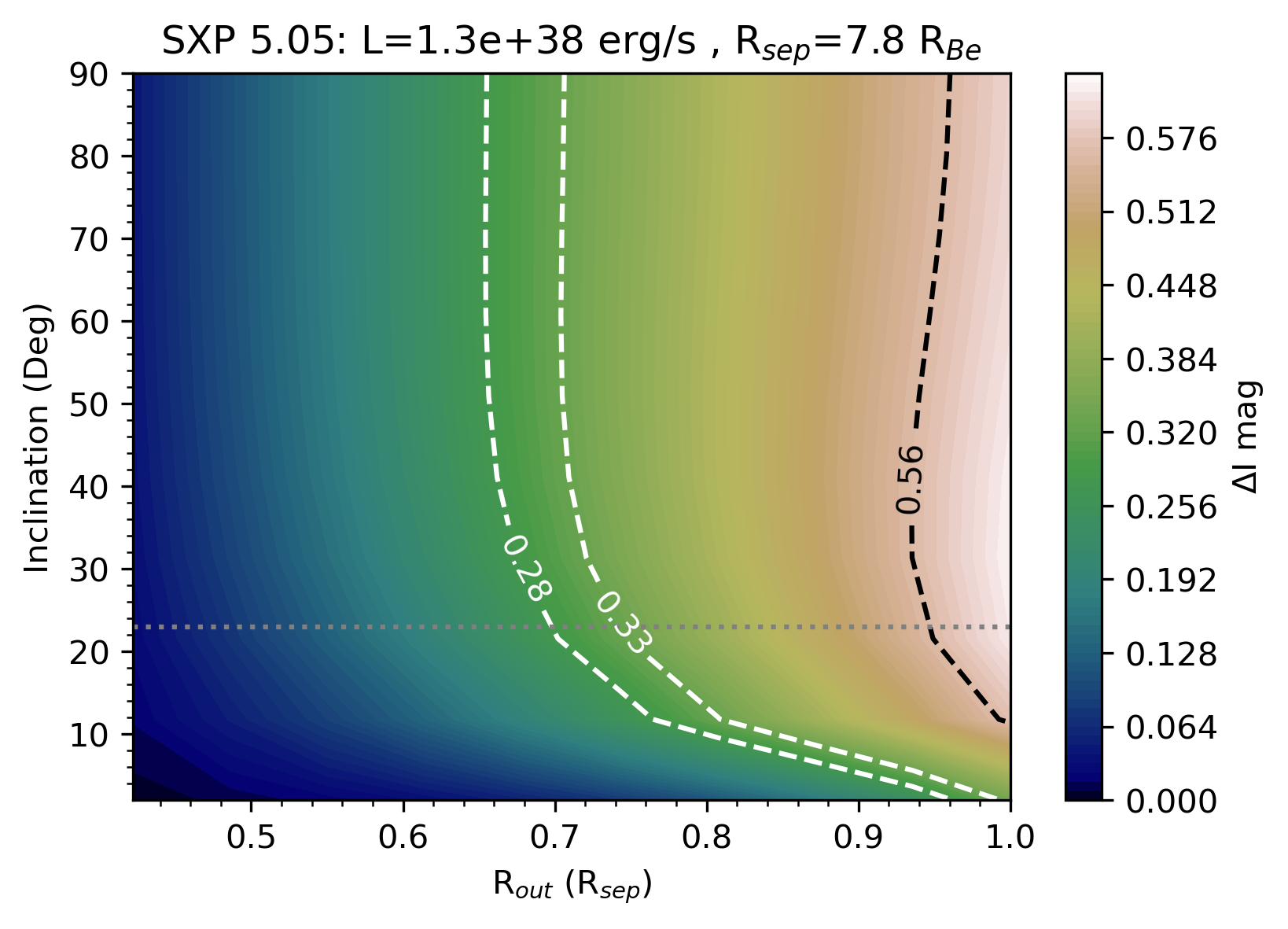}
    \caption{Heat map of simulated maximum (disk is seen face on by observer) flare intensity ($\Delta I$) for SXP\,5.05 for disk inner radius $R_{\rm Be} =3 R_{\star}$. White contours indicate the observed $\Delta I$ (Table \ref{tab:flares}), 
    while the black contour is computed assuming the disk is seen at some moderate inclination  (i.e. 60$^\circ$). Minimum $R_{\rm out}$ is the Be disk size and maximum should be $\sim0.8R_{sep}$ which is roughly the Roche-lobe radius.
    }
    \label{fig:heat1}
\end{figure}

We constructed a two-dimensional grid of configurations for $\theta_{NS}$ and $R_{out}$, and for each grid point, we computed the change in $\Delta I$ compared to the baseline magnitude of the system (i.e. Be star and disk) in the I band. 
In Fig. \ref{fig:heat1}, \ref{fig:heat} \& \ref{fig:heat2} we present heat map plots of the computed $\Delta I$ values and compare them with the observed optical flares. 
When compared to observed $\Delta I$ flare values (see Table \ref{tab:flares}), heat maps may be used to put a limit to the minimum disk size that would be able to reproduce these flares (see white contour lines in heat maps). 
As $\Delta I$ heat maps were created for a face-on observer they represent the maximum change in color for a particular orientation, thus assuming a moderate inclination and correcting for it, the observed flares could be a factor of 2 stronger (see black contour lines). 
For SXP\,5.05 where $\theta_{\rm LOS}\sim83^\circ$, if the Be disk plane has a $\theta_{\rm NS}=23^\circ$ inclination to the orbital plane, then an observer would see the Be disk with $\theta_{\rm LOS}\sim60^\circ$ thus corrected $\Delta I$ would be double than the observed. Therefore we may infer that $R_{\rm out}\sim0.75R_{\rm sep}$ (see horizontal dashed line in Fig. \ref{fig:heat1}) and the Be disk fills the Roche-lobe surface. Note that for heat map in Fig. \ref{fig:heat1} we adopted a larger inner Be disk radius as otherwise the baseline luminosity of the disk is too high to explain the flare intensity (see Fig. \ref{fig:heat1}).

\subsection{$L_{\rm X}$ and flare intensity: observational plane}\label{param_search}

\begin{figure}
    \centering
    \includegraphics[width=\columnwidth]{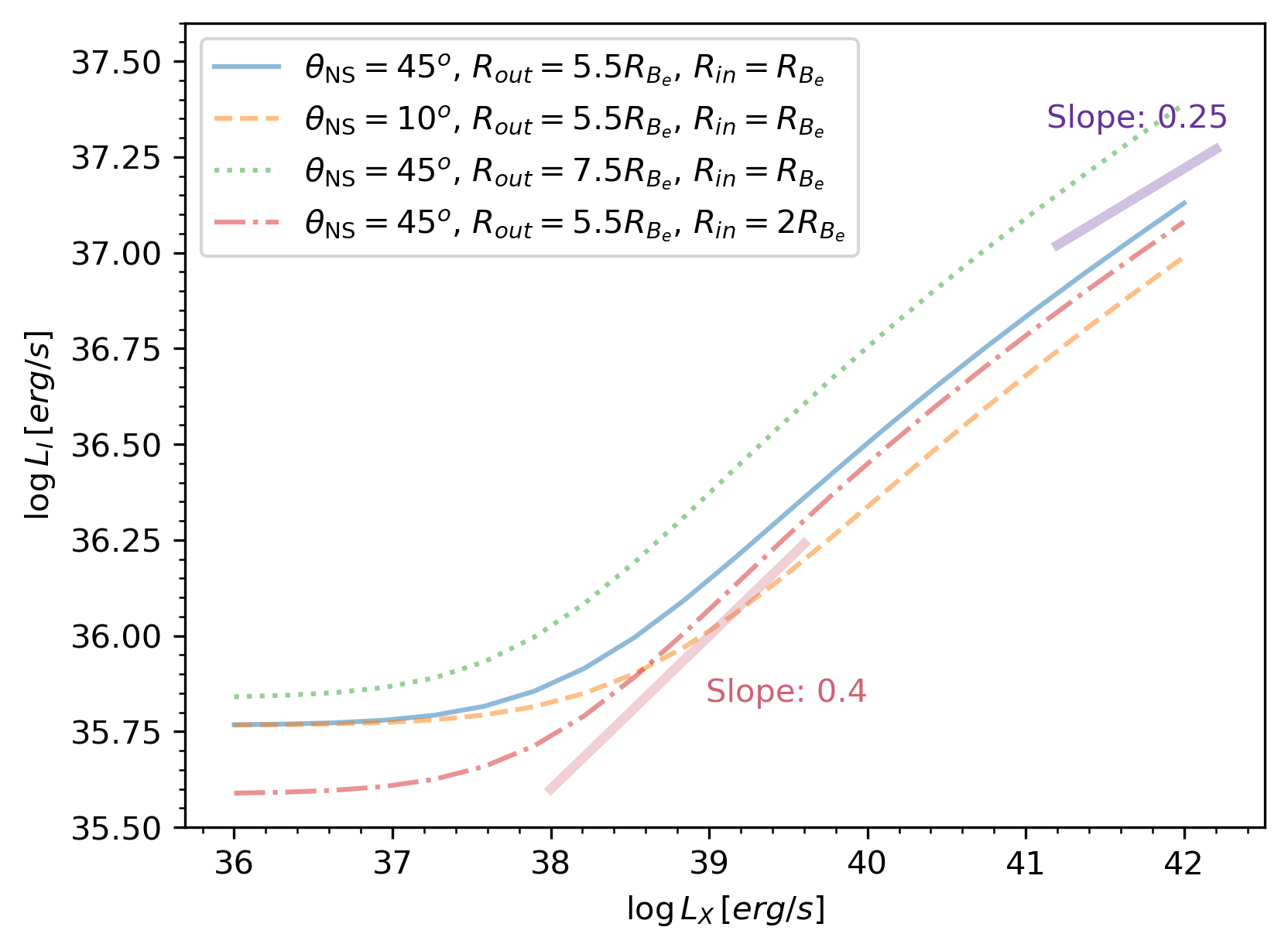}
    \caption{Evolutionary tracks of optical flares luminosity as a function of $L_{\rm X}$ for different model parameters. }
    \label{fig:plane_sim}
\end{figure}

Let us start with a parametric example to compare the luminosity of the irradiated disk as a function of $L_{\rm X}$. We performed simulations for a grid of parameters, computed the SED (similar to  Fig. \ref{fig:sed}), and the luminosity of the optical flares in the I band $L_{\rm I}$ as a function of the X-ray luminosity $L_{\rm X}$. The results are plotted in Fig. \ref{fig:plane_sim}, where we also have scaled all $L_{\rm I}$ by a factor of 0.7, to account for a typical Be disk inclination compared to the observer. For low $L_{\rm X}$ the evolution is flat, indicating that the Be disk is barely heated and maintains its normal temperature. At higher $L_{\rm X}$, between 1-10\ergs{38} we see a linear trend with a slope of $\sim0.3-0.4$ (depending on model parameters). At very high $L_{\rm X}$ the slope is becoming shallower towards a value of 0.25. This is consistent with the I band reaching the Rayleigh-Jeans limit as due to increasing $T_{eff}$ the peak of the irradiated Be disk SED shifts toward lower wavelengths. We also note the presence of a significant scatter based on just altering 2 main configuration parameters $\theta_{\rm NS}$ and $R_{\rm out}$.

\begin{figure}[h]
    \centering
    \includegraphics[width=\columnwidth]{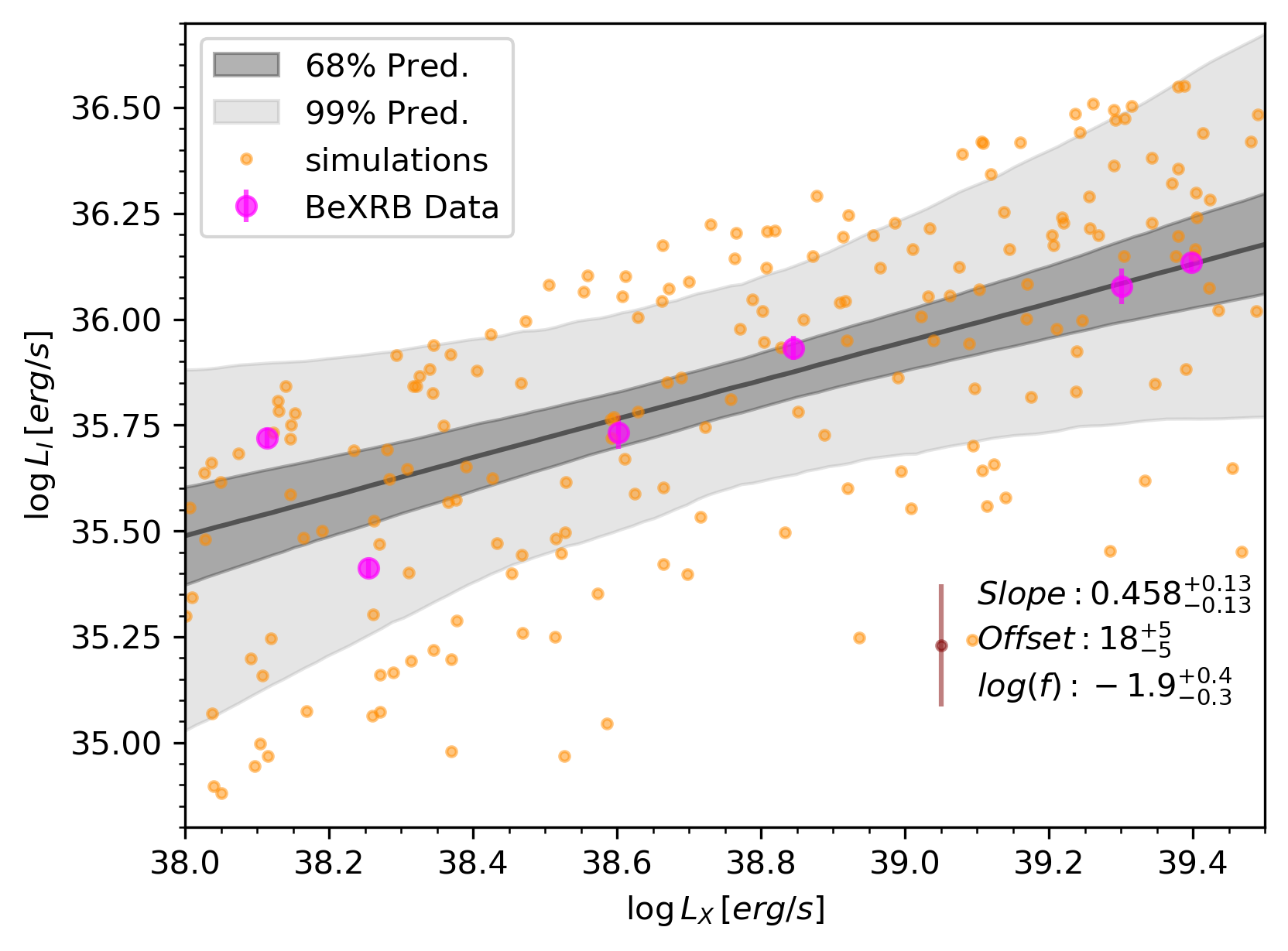}
    \caption{Flare intensity as a function of $log(L_{\rm X})$, where observed data from BeXRBs are plotted with magenta colors. The best-fit line (black) and the 68-99$\%$ prediction bands are plotted with gray shade, while model parameters are given in the legend. For the modeled excess variance we also plot an indicative spread with a maroon color, for easier comparison with the data. Orange points represent simulations for a random set of parameters (see text for details).
    }
    \label{fig:plane_fit}
\end{figure}
 
To compute the observational plane of $L_{\rm X}$ and flare intensity we can start with a comparison of $L_{\rm X}$ and I-band flare amplitude $A$ from table \ref{tab:flares}.
This, however, has several biases since sources have different baseline luminosity levels, thus a $\Delta{\rm mag}$ of 0.2 would translate to a much different flare luminosity. Moreover, sources are at various distances, and suffer from different extinctions, effects that have already been accounted for when computing $L_{\rm X}$. The flare luminosity of each system is then computed as: 
\begin{equation}
L_{\text{flare}} = L_{\rm I} = 4 \pi \text{D}^2 \left( F_{0, I} \cdot 10^{-0.4 \cdot ( \rm I_{BG}- A_{\rm I})} \cdot \left( 10^{0.4 \cdot \Delta I} - 1 \right) \right),
\end{equation}
where $F_{0, I} = 112.6 \times 10^{-11} \text{erg} \, \text{s}^{-1} \, \text{cm}^{-2} \, \text{\AA}^{-1}$ is the reference luminosity for zero mag for the I filter ($\sim7950 \text{\AA}$), and $A_{\rm I}$ the correction for the I-band extinction.
We plotted $L_{\rm I}$ and $L_{\rm X}$ in logarithmic space and proceeded to model them with a linear model including some intrinsic scatter. We used the MLFriends nested sampling MC algorithm implemented via the {\sc ultranest}\footnote{\url{https://johannesbuchner.github.io/UltraNest/}} package \citep{2021JOSS....6.3001B}. We also introduced an excess noise term $\log{f}$ to account for the systematic scatter and noise of our data not included in the statistical uncertainties of the measurements and model. This term is added to the variance of the measurement errors as $\exp(2\log{f})$. The results of the modeling are presented in Fig. \ref{fig:plane_fit}. Interestingly there appears to be a somehow linear relation with a slope $A$ of $\sim$0.46 between two observed quantities and an offset $B$ of 18, which is comparable to the fit onto simulated data obtained in the previous section (see Fig. \ref{fig:plane_sim}).
To further investigate this similarity we performed model simulations for $\theta_{NS}\in[(5^\circ, 90^\circ]$, $\log(L_{\rm X})\in[38, 39.5]$, $R_{out}/R_{B_e}\in[4,9]$ and $R_{in}/R_{B_e}\in[1,2]$, with the results overlaid in Fig. \ref{fig:plane_fit}.

Transforming the linear solution obtained from our fit to observable BeXRB data to an expression into linear space we get:
\begin{equation}
\begin{aligned}
L_{\rm I,35} \approx  3 {L_{\rm X,38}}^{0.46} \rm erg/s
\end{aligned}
\end{equation}
which provides a relation between the normalized $L_{\rm I,35}$ and $L_{\rm X,38}$ luminosities. The numerical factor 3 should be characteristic of I band flares, while a different scaling relation could be extracted for other optical bands.

\section{Discussion}

In this work, we have constructed a numerical model to study the effects of X-ray irradiation in major outbursts of BeXRB systems (see Table \ref{tab:bexrbs}). 
The main free parameters of the model are the geometric configuration of the system and $L_{\rm X}$ originating from the accreting NS.
To investigate the validity of our model assumptions we have compared it with optical flares of six BeXRBs (see Fig. \ref{fig:flares}, and Table \ref{tab:flares}).
A parametric investigation of the model indicates that in the $L_{\rm X}$ range of 1-30\ergs{39} that includes the brighter Type-II outbursts of BeXRBs, the model predicts a linear relation between $L_{\rm X}$ and optical flare intensity. When compared to the observed properties of flares the slope of the model is consistent (within errors) to the observed one, providing hints of a fundamental plane of Type-II outbursts and optical flares. Nevertheless, there is a significant scatter in this relation arising from model parameters like the NS-disk inclination and disk truncation. 
Our approach was also based on a symmetric Be decretion disk, while simulations have shown that in major outbursts the Be disks tend to become elliptic \citep{Martin2011,2021ApJ...922L..37M,2024MNRAS.528L..59M}. Moreover, there is observational evidence of warped disks during outbursts \citep{2018A&A...619A..19R}. However, qualitatively this is not an issue as in most configurations the NS would mainly illuminate one side of the disk (Fig. \ref{fig:Bedisks}), since in our approach we compare the maximum optical intensity for a configuration which would occur at some part of the orbit when the NS illuminates the maximum possible area of the disk.

For individual systems, we can produce multi-parameter contour plots and compare the max difference in I mag during a flare ($\Delta I$) with geometrical configurations (see Fig. \ref{fig:heat1} \& \ref{fig:heat}). 
Interpreting $\Delta I$ heat maps can also be challenging since we do not always know the inclination of the Be disk compared to the observer. 
While constraints may be put from the shape of the $H_{\alpha}$ line, at the same time, there is a degeneracy of predicted $\Delta I$ with other model parameters like $R_{\rm in}$, as for larger $R_{\rm in}$ values the model would produce larger $\Delta I$, as the comparison baseline would be fainter. In cases like SXP\,5.05 where the Be disk size has been constrained from measurements of $H_{\alpha}$ lines \citep[e.g.][]{2015MNRAS.447.2387C,2024MNRAS.528.7115C}, such contours are helpful as they can help constrain the maximum disk size (Fig. \ref{fig:heat1}).
Regardless, there are a lot of caveats and degeneracy that may limit this approach to only qualitative discussions. For example the baseline Be disk temperature profile may play an important role. Following eq. \ref{eq2}, a sharper drop (e.g. $n=2$) in temperature would result in fainter baseline and thus smaller disk sizes would be required to explain an observed $\Delta I$ value. Similar effect would have the change in inner $B_{\rm e}$ disk radius, as it would remove the hotter hart of the disk, again providing a fainter baseline luminosity. 
Finally, another caveat is the adoption of an albedo parameter of 0.5. It is known that soft X-rays heat up the disk surface while hard X-rays penetrate deeper into the disk \citep[e.g.]{2001MNRAS.320..177W,2005MNRAS.362...79C}, thus accounting for the X-ray spectral shape is important. BeXRB pulsars have generally hard spectra with a cut-off and emit more than 40\% of their energy below 10 keV \citep{2022MNRAS.513.1400A}. In our approach we used a single albedo value as a simplification to these effects.
Without multi-wavelength data and accurate radiative transfer implementation it is hard to break degeneracies between the above mentioned parameters. If we focus on SXP\,5.05 a higher albedo would not be possible to explain the flare intensity with a temperature profile provided by eq .\ref{eq3}. A somehow steeper temperature profile and larger inner Be disk radius would be necessary to explain the optical flares. Alternatively one would need to introduce more complex disk geometry and perhaps an eccentric or warped Be disk \citep{2021ApJ...922L..37M}.

\begin{figure}
    \centering
    \includegraphics[width=\columnwidth]{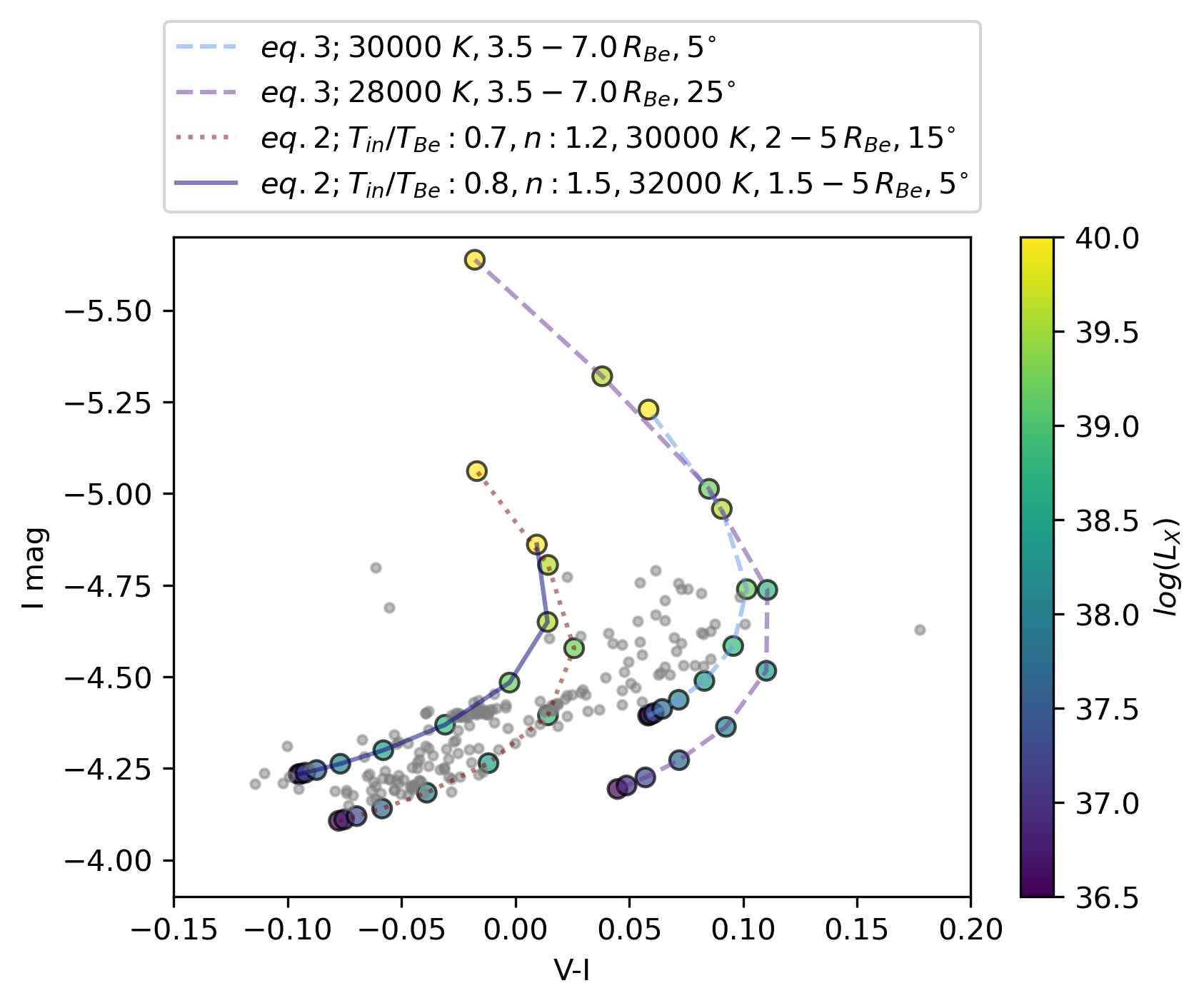}
    \caption{Simulated evolutionary paths for V-I color and absolute systems magnitude of optical flares. On the plot, we overlay data for J0243 corrected for extinction ($A_V=3.2$~mag) and for a distance of 6.3 kpc, values selected to obtain a visual agreement to the simulated curves with temperature profiles from eq. \ref{eq2} and \ref{eq3}.}
    \label{fig:J0243}
\end{figure}

The Be disk irradiation model may also naturally explain the color-magnitude changes in major outbursts of BeXRBs. Study of color-magnitude changes is common for X-ray binaries, where there is strong irradiation of the donor star, and the accretion disk from X-rays emitted close to the compact object. For example studies have been performed for transition between the hard and soft states and its imprint as a non-thermal component seen at optical wavelengths \citep{2020A&A...638A.127K}.  Another example is the effect of irradiation in ULXs and possible how this can induce orbital signatures \citep[e.g.][]{2005MNRAS.362...79C,2006IAUS..230..300C}, where these share similarities with orbital variability we see in this study in the form of optical mini-flares on the OGLE data. 
Although a detailed investigation of these effects on BeXRB systems may be the focus of a follow-up work we would like to offer a qualitative example.
We used the V and I magnitudes from J0243 monitoring \citep{2022A&A...666A.110L} and compared them to simulated color magnitude diagrams. We followed a procedure similar to the parametric modeling that was used to create evolutionary paths shown in Fig. \ref{fig:plane_sim} for 4 sets of parameters. For all sets, we adopted $R_{\rm in}=R_{\rm Be}=8R_{\odot}$, an orbital separation of 12.4$R_{\rm Be}$, and we also assumed that the disk plane is inclined to the line of sight so luminosity of flares is reduced by 50\%. We then experimented with star temperature and inclination $\theta_{\rm NS}$ of the NS orbit compared to the Be disk. In Fig. \ref{fig:J0243} we plot the computed V and I absolute magnitudes together with extinction-corrected absolute magnitudes for J0243. Matching the data with the simulated grid was done by eye and adopting distance and extinction values within an acceptable range. 
We found that for a distance of 6.3 kpc and $A_V=3.3$~mag we get an excellent agreement within observations and data, while the results seem to favor lower $\theta_{\rm NS}$ angles.
Our investigation is partially qualitative as we treat the star as a pure black-body and do not adopt a stellar atmosphere template, this mainly cause the baseline point of the models (lower left in Fig. \ref{fig:J0243}) to shift, however the evolutionary tracks would remain unchanged.
A few general conclusions can also be derived from this investigation. Firstly, the X-ray irradiation model may reproduce the general redder when brighter pattern observed in BeXRBs during outbursts. Also, as X-ray luminosity gets higher, the color changes reach a plateau, and after a critical $L_{\rm X}$ around 1-10\ergs{38} the color of the system becomes bluer. These characteristics are consistent with the behavior of J0243 as reported by \citet{2022A&A...666A.110L}. Interestingly, disks with sharper drops in temperature profile, or inner truncation  ere favored.

Recently, \citet{2024A&A...683A..45A} studied the optical/UV light curves and compared them to the X-ray fluxes along the 2017 outburst of J0243. In their findings, they contribute optical/UV flare to a combination of X-ray irradiation of the Be star surface and the accretion disk. However, they overestimate the heating of the Be Star surface 
(see our Fig. \ref{fig:Bedisks}), while ignoring the contribution of the Be disk irradiation\footnote{The authors are now incorporating an irradiated, misaligned Be disk into their model. Preliminary results (Alfonso-Garzón et al., submitted, erratum) confirm that this component is essential to explain the observations.}. Here, we offer a qualitative explanation for the optical flare in the 2017 outburst of J0243. While detailed modeling of the complete optical/UV light curves is beyond the scope of this work, it demonstrates the importance of X-ray irradiation of the Be disk.

Our findings would perhaps hold higher importance for the study of the population than individual sources. Possible applications would include searching archival optical data to identify similarly bright optical flares which would indirectly probe ``missing'' major Type-II outbursts from BeXRBs in the MCs or even in our Galaxy. Another interesting application would be to investigate if our findings hold at fainter optical flares perhaps associated with Type-I outbursts as there is a plethora of systems that show these flares with $\Delta I \sim 0.02$ and $L_{\rm X}$ between 1-10\ergs{36} \citep{2013A&A...558A..74V,2017MNRAS.470.1971V,2017MNRAS.470.4354V,2014MNRAS.444.3571S}. However, in these systems, the asymmetry of the disk and a variable inclination of the irradiated disk elements as a function of the orbital phase adds to the complexity of the problem.

Finally, it is interesting to explore a possible connection with Ultra-luminous X-ray sources \citep[ULXs, see review][]{2023NewAR..9601672K}. In particular ULXs and ULX pulsars (ULXPs) have shown optical and X-ray periodicities that are not always equal to each other and could be aliases between the binary orbital period and a superorbital period \citep{2021A&A...651A..75F}.
In the X-rays, the super-orbital period could be related to the precession of the accretion disk and the outflows from the accreting sources \citep[e.g.][]{2021ApJ...909...50V,2022MNRAS.509.2493K,2023A&A...672A.140F}.
For the irradiation of stellar companion the binary orbit is much tighter in some ULXPs and $L_{\rm X}$ is close or above \oergs{40} making it an important factor compared to the sources we studied here. In a recent work \citet{2023MNRAS.524.4302K} studied the X-ray, UV, and optical variability in a large sample of ULXs and they concluded that while the subset may show weakly correlated joint variability, other sources appear to display non-linear relationships between the bands.
Given that systems like NGC\,300\,ULX-1 and some transient ULXs/ULXPs have properties consistent with the presence of a OBe companion \citep[e.g.][]{2018MNRAS.476L..45C,2019ApJ...883L..34H,2023NewAR..9601672K} then for the study of their optical properties emission from an irradiated Be disk should be acounted for.  

\section{conclusion}

We have demonstrated how optical flares in major Type-II outbursts may be attributed to X-ray irradiation of the Be decretion disk by the X-ray emission of the NS. Although application to individual systems might be challenging due to multiple model parameters and degeneracies, this model yields a fundamental linear relation between the logarithms of the  X-ray luminosity from the accreting NS and the maximum luminosity of the optical flares with a slope $\sim0.4-0.5$. The observed scatter in the optical vs. X-ray luminosity plane is also naturally explained by the combination of geometrical properties like the inclination of the orbital plane or the observer compared to the Be disk, and the size of the B disk.   \\

\textit{Data Availability. --} 
Optical data for OGLE are publicly available through X-rom project at \url{https://ogle.astrouw.edu.pl/ogle4/xrom/xrom.html} and ASAS-SN data through \url{https://asas-sn.osu.edu/photometry/71ac277e-d7a0-5387-9385-7be7ba02931f}. 
X-ray light-curves are available trough the \swift/XRT data products generator \citep{2007A&A...469..379E,2009MNRAS.397.1177E}  via \url{https://www.swift.ac.uk/user_objects/}.
Other X-ray and optical data were obtained from the literature and from published papers.
The data and code underlying this article will be shared on reasonable request to the corresponding author.
The results of our work are available in a GitHub repository \url{https://github.com/gevas-astro/BeXRIOT}. This includes: (a) the datasets (b) code for visualization of results in python and jupyter notebooks.

\begin{acknowledgements}
I thank the anonymous referee for all their comments and suggestions that helped improve the manuscript.
This work made use of data supplied by the UK Swift Science Data Centre at the University of Leicester.
I also thank Maria Petropoulou for fruitful discussions during this project.
GV acknowledges support from the Hellenic Foundation for Research and Innovation (H.F.R.I.) through the project ASTRAPE (Project ID 7802). 
\end{acknowledgements}

%
%

\bibliographystyle{aa} 
\bibliography{general}

\begin{thebibliography}{65}
\expandafter\ifx\csname natexlab\endcsname\relax\def\natexlab#1{#1}\fi

\bibitem[{{Alfonso-Garz{\'o}n} {et~al.}(2024){Alfonso-Garz{\'o}n}, {van den Eijnden}, {Kuin}, {F{\"u}rst}, {Rouco Escorial}, {Fabregat}, {Reig}, {Mas-Hesse}, {Jenke}, {Malacaria}, \& {Wilson-Hodge}}]{2024A&A...683A..45A}
{Alfonso-Garz{\'o}n}, J., {van den Eijnden}, J., {Kuin}, N.~P.~M., {et~al.} 2024, \aap, 683, A45

\bibitem[{{Ambrosi} {et~al.}(2022){Ambrosi}, {Zampieri}, {Pintore}, \& {Wolter}}]{2022MNRAS.509.4694A}
{Ambrosi}, E., {Zampieri}, L., {Pintore}, F., \& {Wolter}, A. 2022, \mnras, 509, 4694

\bibitem[{{Anastasopoulou} {et~al.}(2022){Anastasopoulou}, {Zezas}, {Steiner}, \& {Reig}}]{2022MNRAS.513.1400A}
{Anastasopoulou}, K., {Zezas}, A., {Steiner}, J.~F., \& {Reig}, P. 2022, \mnras, 513, 1400

\bibitem[{{Bailer-Jones} {et~al.}(2021){Bailer-Jones}, {Rybizki}, {Fouesneau}, {Demleitner}, \& {Andrae}}]{2021AJ....161..147B}
{Bailer-Jones}, C.~A.~L., {Rybizki}, J., {Fouesneau}, M., {Demleitner}, M., \& {Andrae}, R. 2021, \aj, 161, 147

\bibitem[{{Bessell} {et~al.}(1998){Bessell}, {Castelli}, \& {Plez}}]{1998A&A...333..231B}
{Bessell}, M.~S., {Castelli}, F., \& {Plez}, B. 1998, \aap, 333, 231

\bibitem[{{Buchner}(2021)}]{2021JOSS....6.3001B}
{Buchner}, J. 2021, The Journal of Open Source Software, 6, 3001

\bibitem[{{Carciofi} \& {Bjorkman}(2006)}]{2006ApJ...639.1081C}
{Carciofi}, A.~C. \& {Bjorkman}, J.~E. 2006, \apj, 639, 1081

\bibitem[{{Carpano} {et~al.}(2018){Carpano}, {Haberl}, {Maitra}, \& {Vasilopoulos}}]{2018MNRAS.476L..45C}
{Carpano}, S., {Haberl}, F., {Maitra}, C., \& {Vasilopoulos}, G. 2018, \mnras, 476, L45

\bibitem[{{Coe} {et~al.}(2015){Coe}, {Bartlett}, {Bird}, {Haberl}, {Kennea}, {McBride}, {Townsend}, \& {Udalski}}]{2015MNRAS.447.2387C}
{Coe}, M.~J., {Bartlett}, E.~S., {Bird}, A.~J., {et~al.} 2015, \mnras, 447, 2387

\bibitem[{{Coe} {et~al.}(2024){Coe}, {Kennea}, {Monageng}, {Townsend}, {Buckley}, {Williams}, {Udalski}, \& {Evans}}]{2024MNRAS.528.7115C}
{Coe}, M.~J., {Kennea}, J.~A., {Monageng}, I.~M., {et~al.} 2024, \mnras, 528, 7115

\bibitem[{{Coe} {et~al.}(2001){Coe}, {Negueruela}, {Buckley}, {Haigh}, \& {Laycock}}]{2001MNRAS.324..623C}
{Coe}, M.~J., {Negueruela}, I., {Buckley}, D.~A.~H., {Haigh}, N.~J., \& {Laycock}, S.~G.~T. 2001, \mnras, 324, 623

\bibitem[{{Copperwheat} {et~al.}(2005){Copperwheat}, {Cropper}, {Soria}, \& {Wu}}]{2005MNRAS.362...79C}
{Copperwheat}, C., {Cropper}, M., {Soria}, R., \& {Wu}, K. 2005, \mnras, 362, 79

\bibitem[{{Copperwheat} {et~al.}(2006){Copperwheat}, {Cropper}, {Soria}, \& {Wu}}]{2006IAUS..230..300C}
{Copperwheat}, C., {Cropper}, M., {Soria}, R., \& {Wu}, K. 2006, in IAU Symposium, Vol. 230, Populations of High Energy Sources in Galaxies, ed. E.~J.~A. {Meurs} \& G.~{Fabbiano}, 300--301

\bibitem[{{de Wit} {et~al.}(2006){de Wit}, {Lamers}, {Marquette}, \& {Beaulieu}}]{deWit2006}
{de Wit}, W.~J., {Lamers}, H.~J.~G.~L.~M., {Marquette}, J.~B., \& {Beaulieu}, J.~P. 2006, \aap, 456, 1027

\bibitem[{{Evans} {et~al.}(2009){Evans}, {Beardmore}, {Page}, {Osborne}, {O'Brien}, {Willingale}, {Starling}, {Burrows}, {Godet}, {Vetere}, {Racusin}, {Goad}, {Wiersema}, {Angelini}, {Capalbi}, {Chincarini}, {Gehrels}, {Kennea}, {Margutti}, {Morris}, {Mountford}, {Pagani}, {Perri}, {Romano}, \& {Tanvir}}]{2009MNRAS.397.1177E}
{Evans}, P.~A., {Beardmore}, A.~P., {Page}, K.~L., {et~al.} 2009, \mnras, 397, 1177

\bibitem[{{Evans} {et~al.}(2007){Evans}, {Beardmore}, {Page}, {Tyler}, {Osborne}, {Goad}, {O'Brien}, {Vetere}, {Racusin}, {Morris}, {Burrows}, {Capalbi}, {Perri}, {Gehrels}, \& {Romano}}]{2007A&A...469..379E}
{Evans}, P.~A., {Beardmore}, A.~P., {Page}, K.~L., {et~al.} 2007, \aap, 469, 379

\bibitem[{{Fitzpatrick} \& {Massa}(2007)}]{2007ApJ...663..320F}
{Fitzpatrick}, E.~L. \& {Massa}, D. 2007, \apj, 663, 320

\bibitem[{{F{\"u}rst} {et~al.}(2021){F{\"u}rst}, {Walton}, {Heida}, {Bachetti}, {Pinto}, {Middleton}, {Brightman}, {Earnshaw}, {Barret}, {Fabian}, {Kretschmar}, {Pottschmidt}, {Ptak}, {Roberts}, {Stern}, {Webb}, \& {Wilms}}]{2021A&A...651A..75F}
{F{\"u}rst}, F., {Walton}, D.~J., {Heida}, M., {et~al.} 2021, \aap, 651, A75

\bibitem[{{F{\"u}rst} {et~al.}(2023){F{\"u}rst}, {Walton}, {Israel}, {Bachetti}, {Barret}, {Brightman}, {Earnshaw}, {Fabian}, {Heida}, {Imbrogno}, {Middleton}, {Pinto}, {Salvaterra}, {Roberts}, {Rodr{\'\i}guez Castillo}, \& {Webb}}]{2023A&A...672A.140F}
{F{\"u}rst}, F., {Walton}, D.~J., {Israel}, G.~L., {et~al.} 2023, \aap, 672, A140

\bibitem[{{Gandhi} {et~al.}(2010){Gandhi}, {Dhillon}, {Durant}, {Fabian}, {Kubota}, {Makishima}, {Malzac}, {Marsh}, {Miller}, {Shahbaz}, {Spruit}, \& {Casella}}]{2010MNRAS.407.2166G}
{Gandhi}, P., {Dhillon}, V.~S., {Durant}, M., {et~al.} 2010, \mnras, 407, 2166

\bibitem[{{Gehrels} {et~al.}(2004){Gehrels}, {Chincarini}, {Giommi}, {Mason}, {Nousek}, {Wells}, {White}, {Barthelmy}, {Burrows}, {Cominsky}, {Hurley}, {Marshall}, {M{\'e}sz{\'a}ros}, {Roming}, {Angelini}, {Barbier}, {Belloni}, {Campana}, {Caraveo}, {Chester}, {Citterio}, {Cline}, {Cropper}, {Cummings}, {Dean}, {Feigelson}, {Fenimore}, {Frail}, {Fruchter}, {Garmire}, {Gendreau}, {Ghisellini}, {Greiner}, {Hill}, {Hunsberger}, {Krimm}, {Kulkarni}, {Kumar}, {Lebrun}, {Lloyd-Ronning}, {Markwardt}, {Mattson}, {Mushotzky}, {Norris}, {Osborne}, {Paczynski}, {Palmer}, {Park}, {Parsons}, {Paul}, {Rees}, {Reynolds}, {Rhoads}, {Sasseen}, {Schaefer}, {Short}, {Smale}, {Smith}, {Stella}, {Tagliaferri}, {Takahashi}, {Tashiro}, {Townsley}, {Tueller}, {Turner}, {Vietri}, {Voges}, {Ward}, {Willingale}, {Zerbi}, \& {Zhang}}]{2004ApJ...611.1005G}
{Gehrels}, N., {Chincarini}, G., {Giommi}, P., {et~al.} 2004, \apj, 611, 1005

\bibitem[{{Guillot} {et~al.}(2018){Guillot}, {Vasilopoulos}, {Pasham}, {Jaisawal}, {Ray}, {Wolff}, {Gendreau}, {Strohmayer}, {Arzoumanian}, {Corcoran}, {Altamirano}, {Wilson-Hodge}, {Antoniou}, {Zezas}, \& {Haberl}}]{2018ATel12219....1G}
{Guillot}, S., {Vasilopoulos}, G., {Pasham}, D., {et~al.} 2018, The Astronomer's Telegram, 12219, 1

\bibitem[{{Haberl} {et~al.}(2017){Haberl}, {Israel}, {Rodriguez Castillo}, {Vasilopoulos}, {Delvaux}, {De Luca}, {Carpano}, {Esposito}, {Novara}, {Salvaterra}, {Tiengo}, {D'Agostino}, \& {Udalski}}]{2017A&A...598A..69H}
{Haberl}, F., {Israel}, G.~L., {Rodriguez Castillo}, G.~A., {et~al.} 2017, \aap, 598, A69

\bibitem[{{Harmanec} {et~al.}(1982){Harmanec}, {Horn}, \& {Koubsky}}]{Harmanec1982}
{Harmanec}, P., {Horn}, J., \& {Koubsky}, P. 1982, in Be Stars, ed. M.~{Jaschek} \& H.~G. {Groth}, Vol.~98, 269

\bibitem[{{Harrison} {et~al.}(2013){Harrison}, {Craig}, {Christensen}, {Hailey}, {Zhang}, {Boggs}, {Stern}, {Cook}, {Forster}, {Giommi}, {Grefenstette}, {Kim}, {Kitaguchi}, {Koglin}, {Madsen}, {Mao}, {Miyasaka}, {Mori}, {Perri}, {Pivovaroff}, {Puccetti}, {Rana}, {Westergaard}, {Willis}, {Zoglauer}, {An}, {Bachetti}, {Barri{\`e}re}, {Bellm}, {Bhalerao}, {Brejnholt}, {Fuerst}, {Liebe}, {Markwardt}, {Nynka}, {Vogel}, {Walton}, {Wik}, {Alexander}, {Cominsky}, {Hornschemeier}, {Hornstrup}, {Kaspi}, {Madejski}, {Matt}, {Molendi}, {Smith}, {Tomsick}, {Ajello}, {Ballantyne}, {Balokovi{\'c}}, {Barret}, {Bauer}, {Blandford}, {Brandt}, {Brenneman}, {Chiang}, {Chakrabarty}, {Chenevez}, {Comastri}, {Dufour}, {Elvis}, {Fabian}, {Farrah}, {Fryer}, {Gotthelf}, {Grindlay}, {Helfand}, {Krivonos}, {Meier}, {Miller}, {Natalucci}, {Ogle}, {Ofek}, {Ptak}, {Reynolds}, {Rigby}, {Tagliaferri}, {Thorsett}, {Treister}, \& {Urry}}]{2013ApJ...770..103H}
{Harrison}, F.~A., {Craig}, W.~W., {Christensen}, F.~E., {et~al.} 2013, \apj, 770, 103

\bibitem[{{Heida} {et~al.}(2019){Heida}, {Lau}, {Davies}, {Brightman}, {F{\"u}rst}, {Grefenstette}, {Kennea}, {Tramper}, {Walton}, \& {Harrison}}]{2019ApJ...883L..34H}
{Heida}, M., {Lau}, R.~M., {Davies}, B., {et~al.} 2019, \apjl, 883, L34

\bibitem[{{Jaisawal} {et~al.}(2023){Jaisawal}, {Vasilopoulos}, {Naik}, {Maitra}, {Malacaria}, {Chhotaray}, {Gendreau}, {Guillot}, {Ng}, \& {Sanna}}]{2023MNRAS.521.3951J}
{Jaisawal}, G.~K., {Vasilopoulos}, G., {Naik}, S., {et~al.} 2023, \mnras, 521, 3951

\bibitem[{{Karaferias} {et~al.}(2023){Karaferias}, {Vasilopoulos}, {Petropoulou}, {Jenke}, {Wilson-Hodge}, \& {Malacaria}}]{2023MNRAS.520..281K}
{Karaferias}, A.~S., {Vasilopoulos}, G., {Petropoulou}, M., {et~al.} 2023, \mnras, 520, 281

\bibitem[{{Khan} \& {Middleton}(2023)}]{2023MNRAS.524.4302K}
{Khan}, N. \& {Middleton}, M.~J. 2023, \mnras, 524, 4302

\bibitem[{{Khan} {et~al.}(2022){Khan}, {Middleton}, {Wiktorowicz}, {Dauser}, {Roberts}, \& {Wilms}}]{2022MNRAS.509.2493K}
{Khan}, N., {Middleton}, M.~J., {Wiktorowicz}, G., {et~al.} 2022, \mnras, 509, 2493

\bibitem[{{King} {et~al.}(2023){King}, {Lasota}, \& {Middleton}}]{2023NewAR..9601672K}
{King}, A., {Lasota}, J.-P., \& {Middleton}, M. 2023, \nar, 96, 101672

\bibitem[{{Kosenkov} {et~al.}(2020){Kosenkov}, {Veledina}, {Suleimanov}, \& {Poutanen}}]{2020A&A...638A.127K}
{Kosenkov}, I.~A., {Veledina}, A., {Suleimanov}, V.~F., \& {Poutanen}, J. 2020, \aap, 638, A127

\bibitem[{{Kourniotis} {et~al.}(2014){Kourniotis}, {Bonanos}, {Soszy{\'n}ski}, {Poleski}, {Krikelis}, {Udalski}, {Szyma{\'n}ski}, {Kubiak}, {Pietrzy{\'n}ski}, {Wyrzykowski}, {Ulaczyk}, {Koz{\l}owski}, \& {Pietrukowicz}}]{Kourniotis2014}
{Kourniotis}, M., {Bonanos}, A.~Z., {Soszy{\'n}ski}, I., {et~al.} 2014, \aap, 562, A125

\bibitem[{{Liu} {et~al.}(2022){Liu}, {Yan}, {Reig}, {Wang}, {Xiao}, {Lin}, {Zhang}, {Sai}, {Chen}, {Yan}, \& {Liu}}]{2022A&A...666A.110L}
{Liu}, W., {Yan}, J., {Reig}, P., {et~al.} 2022, \aap, 666, A110

\bibitem[{{Maravelias} {et~al.}(2018){Maravelias}, {Antoniou}, {Zezas}, {Strantzalis}, {Hatzidimitriou}, \& {Haberl}}]{2018ATel12224....1M}
{Maravelias}, G., {Antoniou}, V., {Zezas}, A., {et~al.} 2018, The Astronomer's Telegram, 12224, 1

\bibitem[{{Martin}(2023)}]{Martin2023}
{Martin}, R.~G. 2023, \mnras, 523, L75

\bibitem[{{Martin} \& {Charles}(2024)}]{2024MNRAS.528L..59M}
{Martin}, R.~G. \& {Charles}, P.~A. 2024, \mnras, 528, L59

\bibitem[{{Martin} \& {Franchini}(2021)}]{2021ApJ...922L..37M}
{Martin}, R.~G. \& {Franchini}, A. 2021, \apjl, 922, L37

\bibitem[{{Martin} {et~al.}(2011){Martin}, {Pringle}, {Tout}, \& {Lubow}}]{Martin2011}
{Martin}, R.~G., {Pringle}, J.~E., {Tout}, C.~A., \& {Lubow}, S.~H. 2011, \mnras, 416, 2827

\bibitem[{{McBride} {et~al.}(2008){McBride}, {Coe}, {Negueruela}, {Schurch}, \& {McGowan}}]{2008MNRAS.388.1198M}
{McBride}, V.~A., {Coe}, M.~J., {Negueruela}, I., {Schurch}, M.~P.~E., \& {McGowan}, K.~E. 2008, \mnras, 388, 1198

\bibitem[{{Monageng} {et~al.}(2019){Monageng}, {Coe}, {Townsend}, {Buckley}, {McBride}, {Roche}, {Kennea}, {Udalski}, \& {Evans}}]{2019MNRAS.485.4617M}
{Monageng}, I.~M., {Coe}, M.~J., {Townsend}, L.~J., {et~al.} 2019, \mnras, 485, 4617

\bibitem[{{Pecaut} \& {Mamajek}(2013)}]{2013ApJS..208....9P}
{Pecaut}, M.~J. \& {Mamajek}, E.~E. 2013, \apjs, 208, 9

\bibitem[{{Petropoulou} {et~al.}(2018){Petropoulou}, {Vasilopoulos}, {Christie}, {Giannios}, \& {Coe}}]{2018MNRAS.474L..22P}
{Petropoulou}, M., {Vasilopoulos}, G., {Christie}, I.~M., {Giannios}, D., \& {Coe}, M.~J. 2018, \mnras, 474, L22

\bibitem[{{Reig}(2011)}]{Reig2011}
{Reig}, P. 2011, \apss, 332, 1

\bibitem[{{Reig} \& {Blinov}(2018)}]{2018A&A...619A..19R}
{Reig}, P. \& {Blinov}, D. 2018, \aap, 619, A19

\bibitem[{{Reig} {et~al.}(2020){Reig}, {Fabregat}, \& {Alfonso-Garz{\'o}n}}]{2020A&A...640A..35R}
{Reig}, P., {Fabregat}, J., \& {Alfonso-Garz{\'o}n}, J. 2020, \aap, 640, A35

\bibitem[{{Skowron} {et~al.}(2021){Skowron}, {Skowron}, {Udalski}, {Szyma{\'n}ski}, {Soszy{\'n}ski}, {Wyrzykowski}, {Ulaczyk}, {Poleski}, {Koz{\l}owski}, {Pietrukowicz}, {Mr{\'o}z}, {Rybicki}, {Iwanek}, {Wrona}, \& {Gromadzki}}]{2021ApJS..252...23S}
{Skowron}, D.~M., {Skowron}, J., {Udalski}, A., {et~al.} 2021, \apjs, 252, 23

\bibitem[{{Sturm} {et~al.}(2014){Sturm}, {Haberl}, {Vasilopoulos}, {Bartlett}, {Maggi}, {Rau}, {Greiner}, \& {Udalski}}]{2014MNRAS.444.3571S}
{Sturm}, R., {Haberl}, F., {Vasilopoulos}, G., {et~al.} 2014, \mnras, 444, 3571

\bibitem[{{Tendulkar} {et~al.}(2014){Tendulkar}, {F{\"u}rst}, {Pottschmidt}, {Bachetti}, {Bhalerao}, {Boggs}, {Christensen}, {Craig}, {Hailey}, {Harrison}, {Stern}, {Tomsick}, {Walton}, \& {Zhang}}]{2014ApJ...795..154T}
{Tendulkar}, S.~P., {F{\"u}rst}, F., {Pottschmidt}, K., {et~al.} 2014, \apj, 795, 154

\bibitem[{{Townsend} {et~al.}(2011){Townsend}, {Coe}, {Corbet}, \& {Hill}}]{2011MNRAS.416.1556T}
{Townsend}, L.~J., {Coe}, M.~J., {Corbet}, R.~H.~D., \& {Hill}, A.~B. 2011, \mnras, 416, 1556

\bibitem[{{Treiber} {et~al.}(2021){Treiber}, {Vasilopoulos}, {Bailyn}, {Haberl}, {Gendreau}, {Ray}, {Maitra}, {Maggi}, {Jaisawal}, {Udalski}, {Wilms}, {Monageng}, {Buckley}, {K{\"o}nig}, \& {Carpano}}]{Treiber2021}
{Treiber}, H., {Vasilopoulos}, G., {Bailyn}, C.~D., {et~al.} 2021, \mnras, 503, 6187

\bibitem[{{Treiber} {et~al.}(2025){Treiber}, {Vasilopoulos}, {Bailyn}, {Haberl}, \& {Udalski}}]{2025A&A...694A..43T}
{Treiber}, H., {Vasilopoulos}, G., {Bailyn}, C.~D., {Haberl}, F., \& {Udalski}, A. 2025, \aap, 694, A43

\bibitem[{{Tsygankov} {et~al.}(2017){Tsygankov}, {Doroshenko}, {Lutovinov}, {Mushtukov}, \& {Poutanen}}]{2017A&A...605A..39T}
{Tsygankov}, S.~S., {Doroshenko}, V., {Lutovinov}, A.~A., {Mushtukov}, A.~A., \& {Poutanen}, J. 2017, \aap, 605, A39

\bibitem[{{Udalski} {et~al.}(1992){Udalski}, {Szymanski}, {Kaluzny}, {Kubiak}, \& {Mateo}}]{1992AcA....42..253U}
{Udalski}, A., {Szymanski}, M., {Kaluzny}, J., {Kubiak}, M., \& {Mateo}, M. 1992, \actaa, 42, 253

\bibitem[{{Udalski} {et~al.}(2008){Udalski}, {Szymanski}, {Soszynski}, \& {Poleski}}]{2008AcA....58...69U}
{Udalski}, A., {Szymanski}, M.~K., {Soszynski}, I., \& {Poleski}, R. 2008, \actaa, 58, 69

\bibitem[{{Udalski} {et~al.}(2015){Udalski}, {Szyma{\'n}ski}, \& {Szyma{\'n}ski}}]{Udalski2015}
{Udalski}, A., {Szyma{\'n}ski}, M.~K., \& {Szyma{\'n}ski}, G. 2015, \actaa, 65, 1

\bibitem[{{Vasilopoulos} {et~al.}(2017{\natexlab{a}}){Vasilopoulos}, {Haberl}, \& {Maggi}}]{2017MNRAS.470.1971V}
{Vasilopoulos}, G., {Haberl}, F., \& {Maggi}, P. 2017{\natexlab{a}}, \mnras, 470, 1971

\bibitem[{{Vasilopoulos} {et~al.}(2014){Vasilopoulos}, {Haberl}, {Sturm}, {Maggi}, \& {Udalski}}]{2014A&A...567A.129V}
{Vasilopoulos}, G., {Haberl}, F., {Sturm}, R., {Maggi}, P., \& {Udalski}, A. 2014, \aap, 567, A129

\bibitem[{{Vasilopoulos} {et~al.}(2021){Vasilopoulos}, {Koliopanos}, {Haberl}, {Treiber}, {Brightman}, {Earnshaw}, \& {G{\'u}rpide}}]{2021ApJ...909...50V}
{Vasilopoulos}, G., {Koliopanos}, F., {Haberl}, F., {et~al.} 2021, \apj, 909, 50

\bibitem[{{Vasilopoulos} {et~al.}(2013){Vasilopoulos}, {Maggi}, {Haberl}, {Sturm}, {Pietsch}, {Bartlett}, \& {Coe}}]{2013A&A...558A..74V}
{Vasilopoulos}, G., {Maggi}, P., {Haberl}, F., {et~al.} 2013, \aap, 558, A74

\bibitem[{{Vasilopoulos} {et~al.}(2020){Vasilopoulos}, {Ray}, {Gendreau}, {Jenke}, {Jaisawal}, {Wilson-Hodge}, {Strohmayer}, {Altamirano}, {Iwakiri}, {Wolff}, {Guillot}, {Malacaria}, \& {Stevens}}]{2020MNRAS.494.5350V}
{Vasilopoulos}, G., {Ray}, P.~S., {Gendreau}, K.~C., {et~al.} 2020, \mnras, 494, 5350

\bibitem[{{Vasilopoulos} {et~al.}(2017{\natexlab{b}}){Vasilopoulos}, {Zezas}, {Antoniou}, \& {Haberl}}]{2017MNRAS.470.4354V}
{Vasilopoulos}, G., {Zezas}, A., {Antoniou}, V., \& {Haberl}, F. 2017{\natexlab{b}}, \mnras, 470, 4354

\bibitem[{{Vincentelli} {et~al.}(2023){Vincentelli}, {Neilsen}, {Tetarenko}, {Cavecchi}, {Castro Segura}, {del Palacio}, {van den Eijnden}, {Vasilopoulos}, {Altamirano}, {Armas Padilla}, {Bailyn}, {Belloni}, {Buisson}, {C{\'u}neo}, {Degenaar}, {Knigge}, {Long}, {Jim{\'e}nez-Ibarra}, {Milburn}, {Mu{\~n}oz Darias}, {{\"O}zbey Arabac{\i}}, {Remillard}, \& {Russell}}]{2023Natur.615...45V}
{Vincentelli}, F.~M., {Neilsen}, J., {Tetarenko}, A.~J., {et~al.} 2023, \nat, 615, 45

\bibitem[{{Wilson-Hodge} {et~al.}(2018){Wilson-Hodge}, {Malacaria}, {Jenke}, {Jaisawal}, {Kerr}, {Wolff}, {Arzoumanian}, {Chakrabarty}, {Doty}, {Gendreau}, {Guillot}, {Ho}, {LaMarr}, {Markwardt}, {{\"O}zel}, {Prigozhin}, {Ray}, {Ramos-Lerate}, {Remillard}, {Strohmayer}, {Vezie}, {Wood}, \& {NICER Science Team}}]{2018ApJ...863....9W}
{Wilson-Hodge}, C.~A., {Malacaria}, C., {Jenke}, P.~A., {et~al.} 2018, \apj, 863, 9

\bibitem[{{Wu} {et~al.}(2001){Wu}, {Soria}, {Hunstead}, \& {Johnston}}]{2001MNRAS.320..177W}
{Wu}, K., {Soria}, R., {Hunstead}, R.~W., \& {Johnston}, H.~M. 2001, \mnras, 320, 177

\end{thebibliography}

\begin{appendix}
\section{Individual BeXRB flares: model Contour plots}\label{sec:app}

\begin{figure*}
    \centering
    \includegraphics[width=\columnwidth]{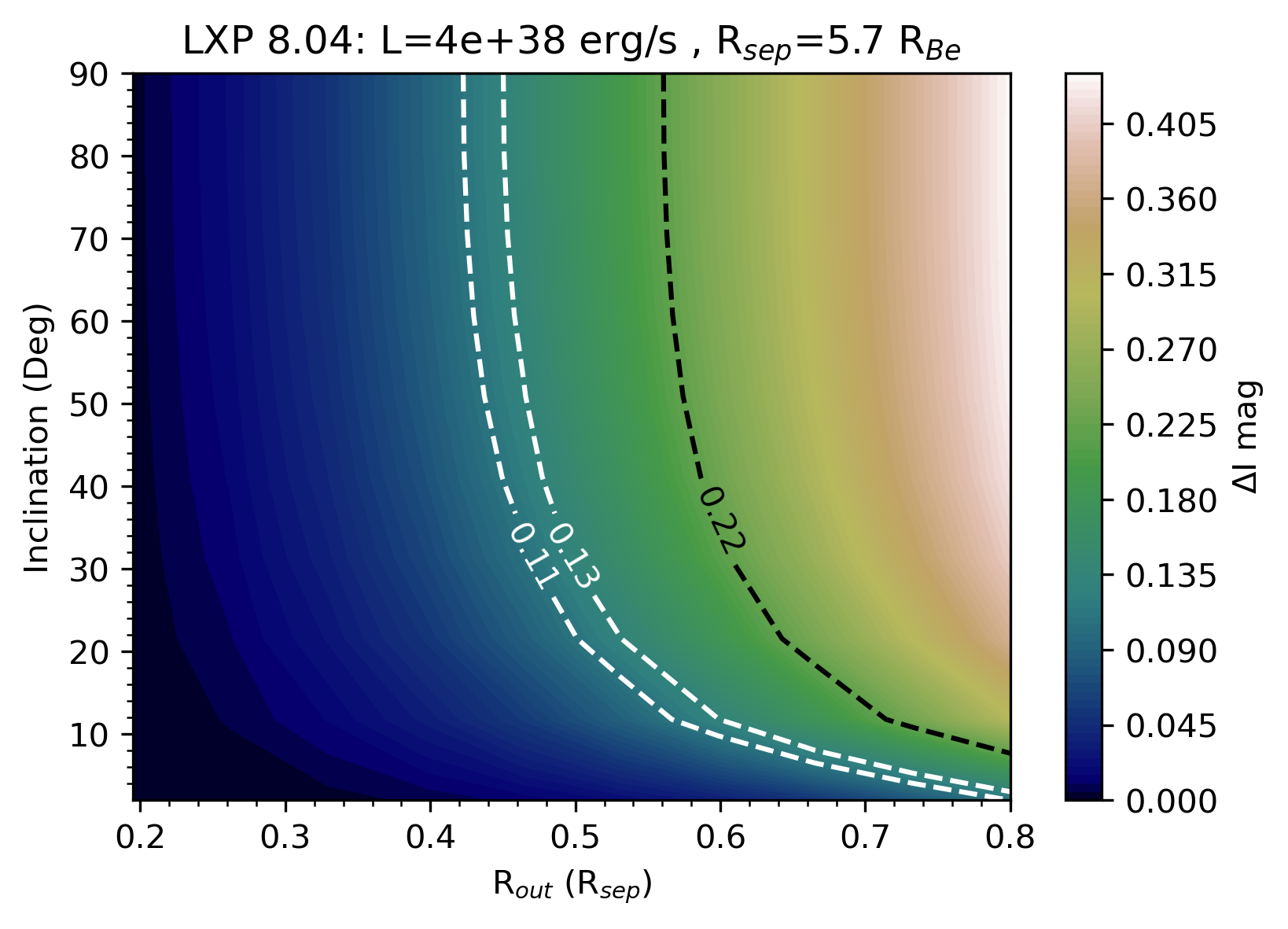}
    \includegraphics[width=\columnwidth]{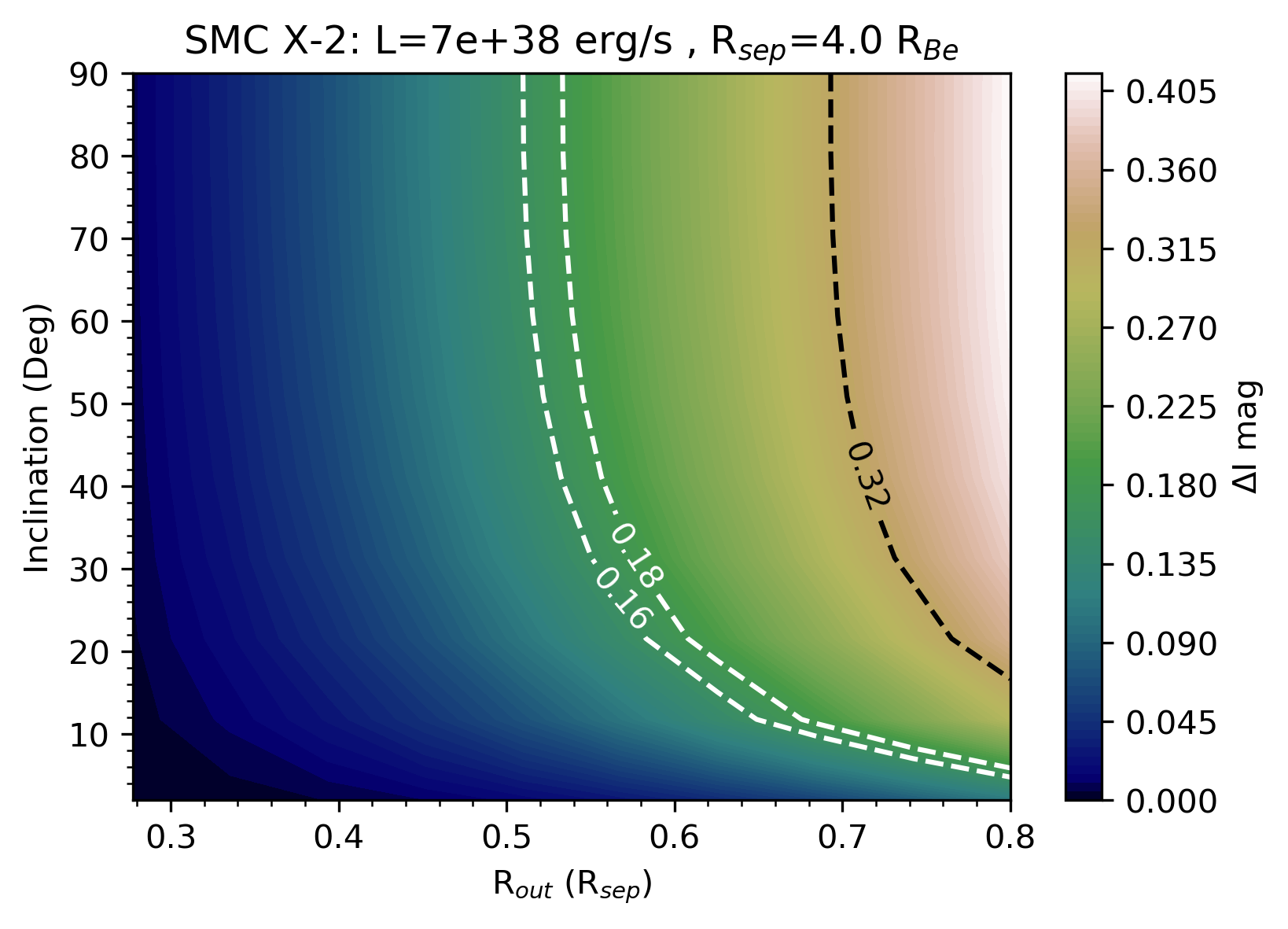}\\
     \includegraphics[width=\columnwidth]{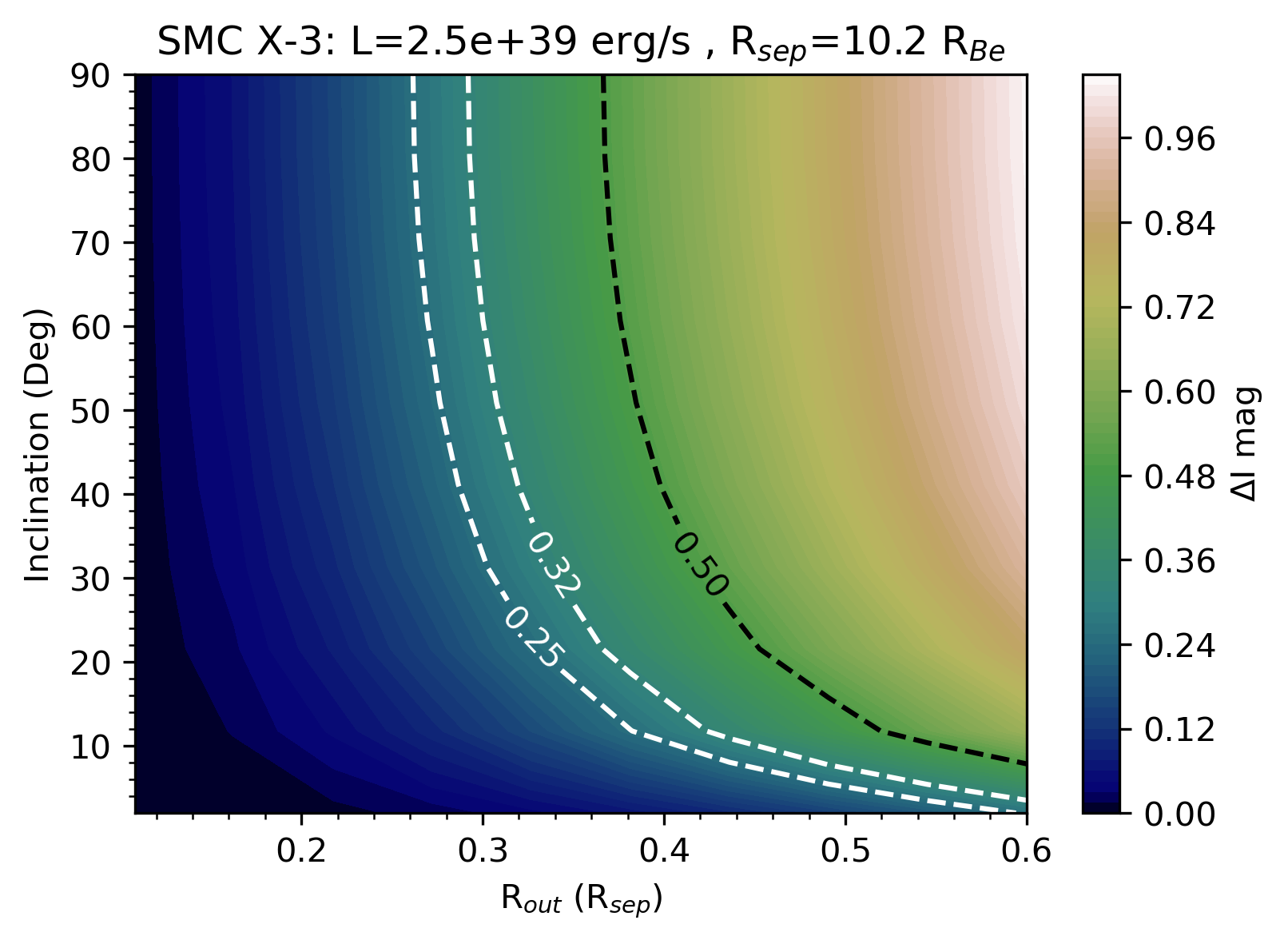}
      \includegraphics[width=\columnwidth]{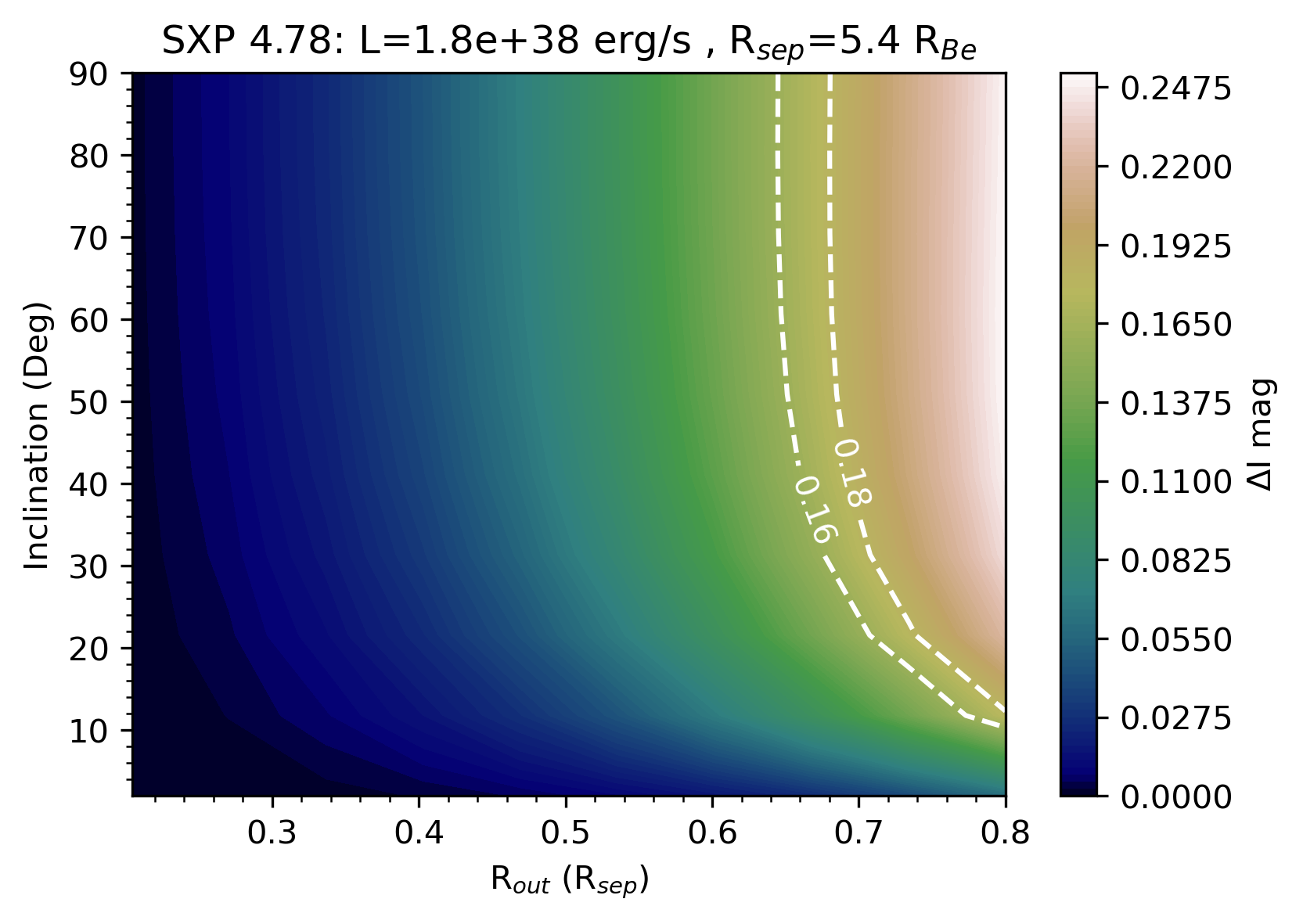}\\
      \includegraphics[width=\columnwidth]{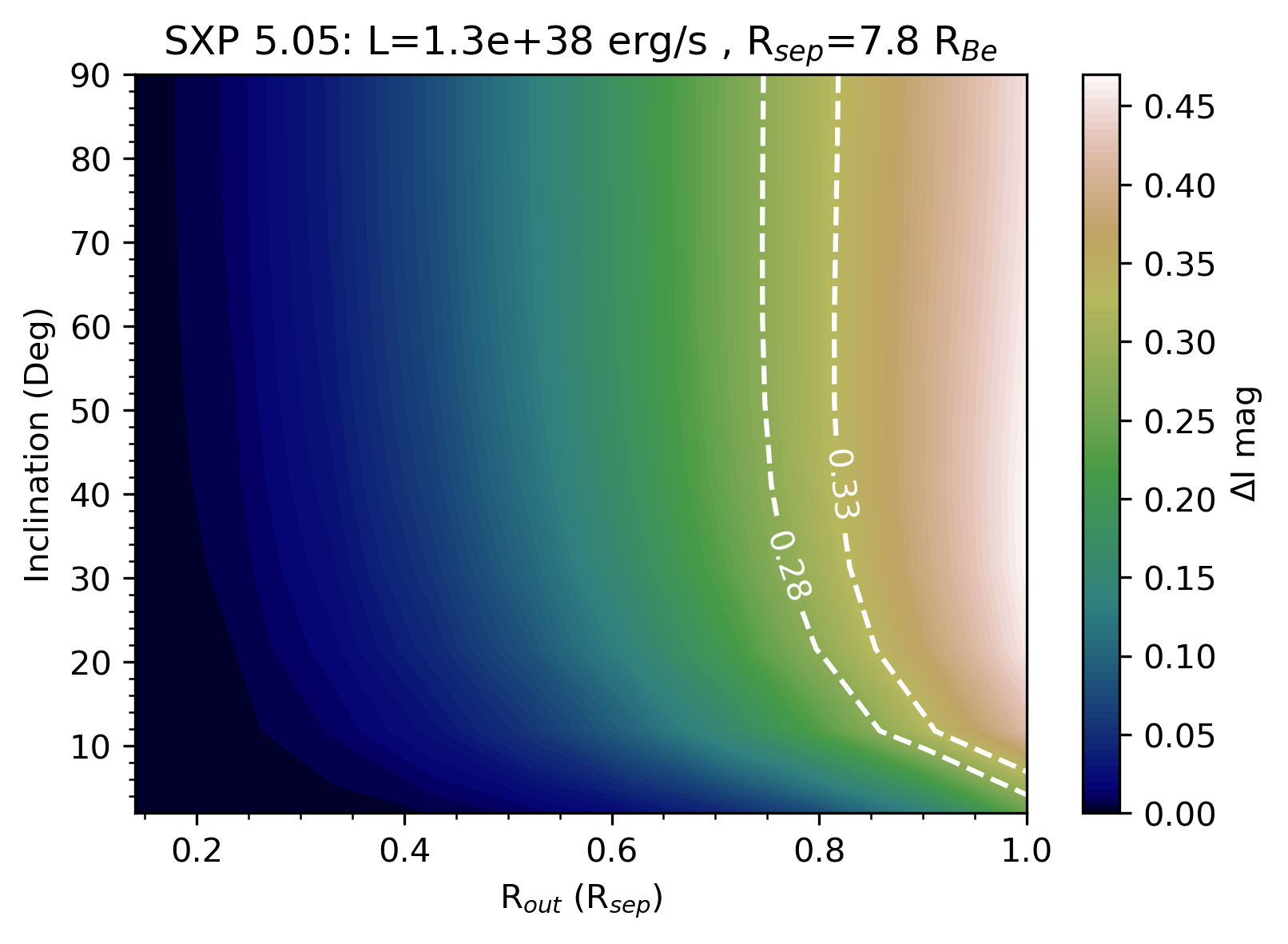}
      \includegraphics[width=\columnwidth]{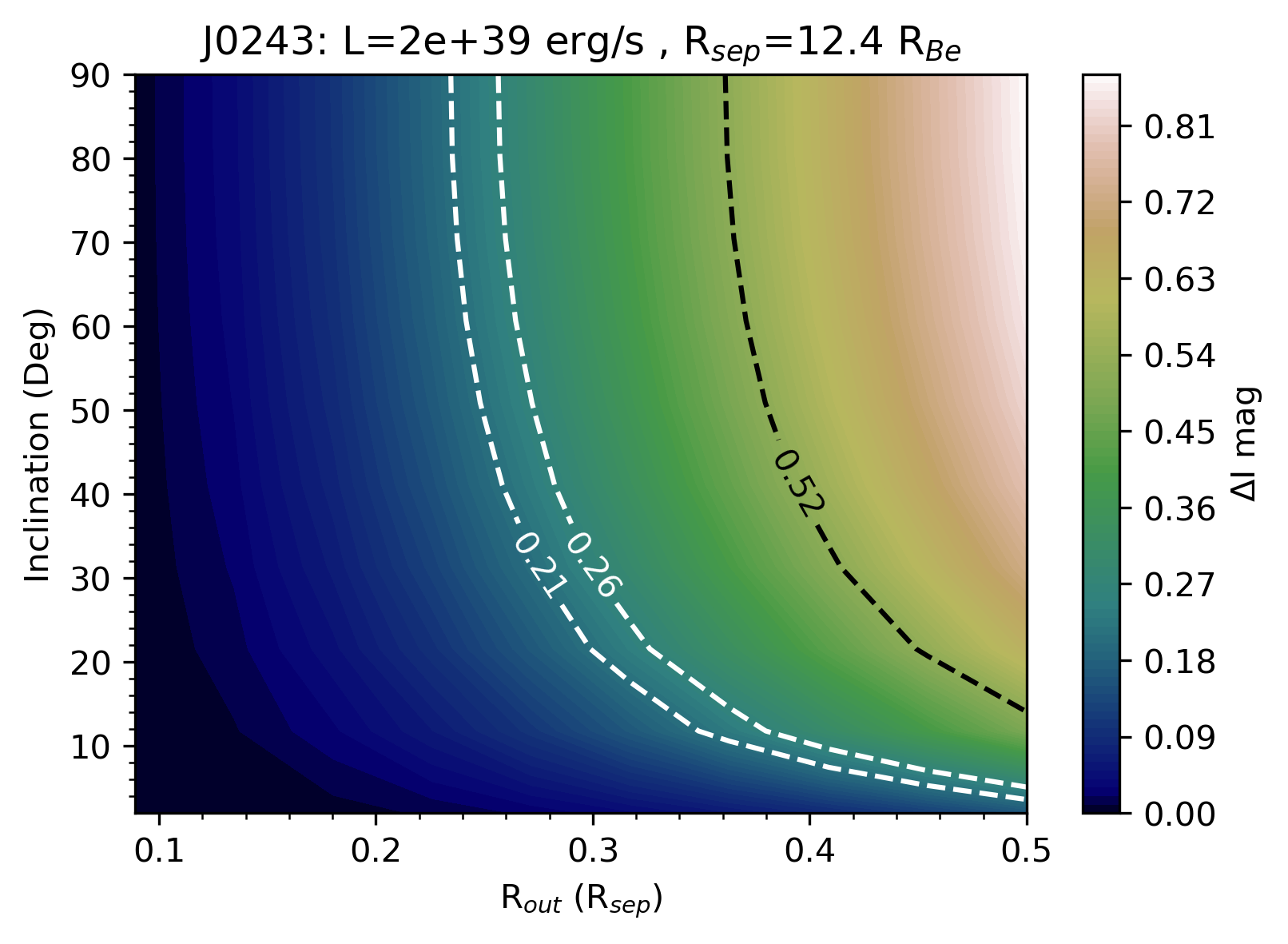}
    \caption{Heat map of simulated maximum (disk is seen face on by observer) flare intensity ($\Delta I$) for parameters of BeXRBs of Table \ref{tab:bexrbs} (see text for details), with a temperature profile of eq. \ref{eq3} and inner disk radius equal to the Be star radius. White contours indicate the observed $\Delta I$ (Table \ref{tab:flares}), while black contour are corrected assuming the disk is seen by some moderate inclination  (i.e. 60$^\circ$). Minimum $R_{\rm out}$ is the Be disk size and maximum should be $\sim0.8R_{sep}$ which is roughly the Roche-lobe.
    }
    \label{fig:heat}
\end{figure*}

Here we present heat maps (Fig. \ref{fig:heat}) for all remaining BeXRB flares. For our calculations we adopt 30000$^o$K and  $8$R$_{\odot}$ for all systems.
For binary separation reasonable separation values would be between 1-2$a\sin{i}$ if an orbital solution is available. For J0243 to be consistent with the literature we adopt 2$a\sin{i}$ \citep{2020A&A...640A..35R}. For other systems with known orbital solutions we adopt an $a\sin{i}$ while using the literature values from Table \ref{tab:bexrbs}, while for SXP\,4.78 we will use an conservative separation of 100 ls.
For comparison we also perform the same heat maps using a temperature profile described in eq. \ref{eq2}, with $n=2$, $T_{\rm disk,0}=T_\star$.

\begin{figure*}
    \centering
    \includegraphics[width=\columnwidth]{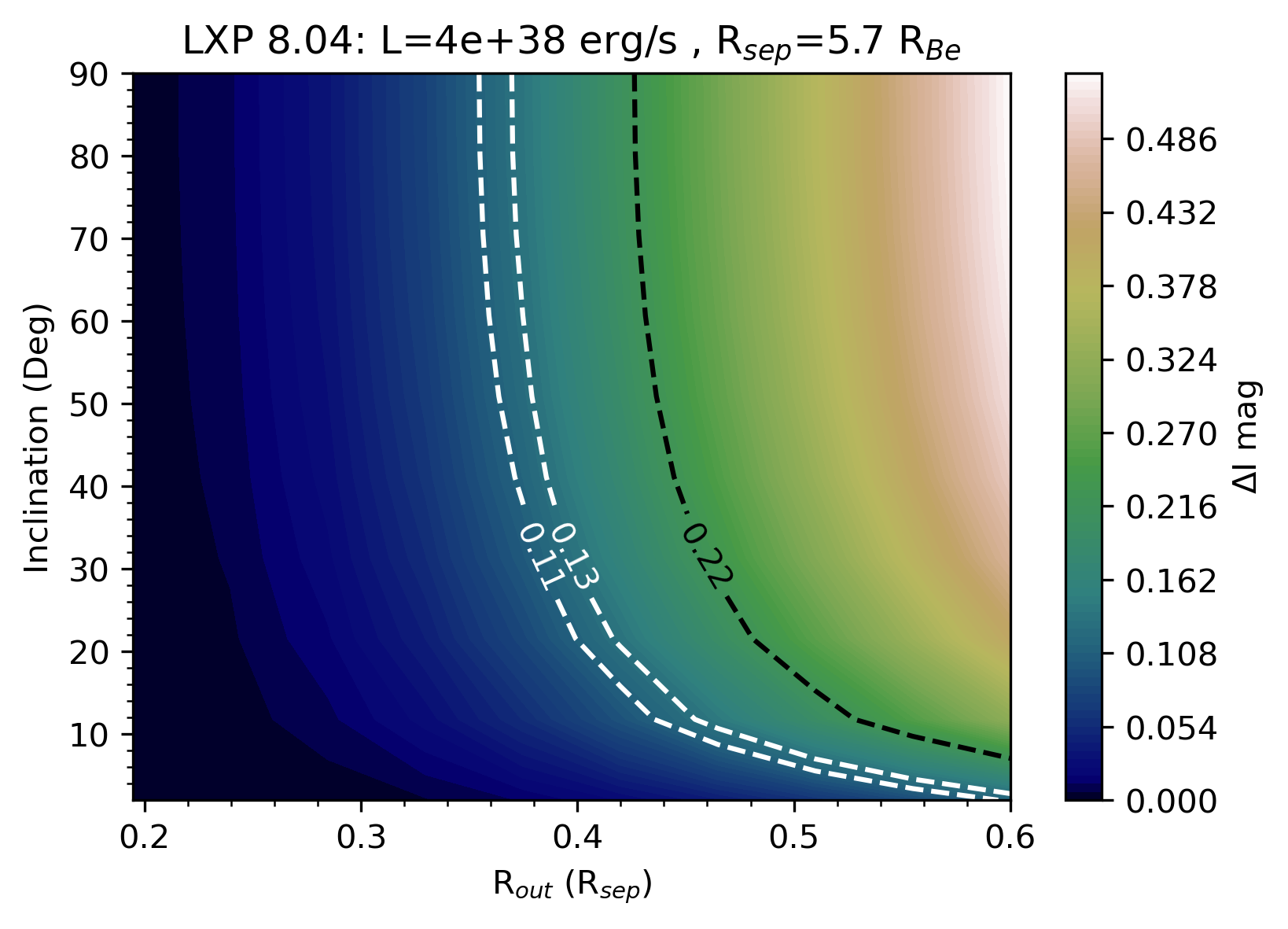}
    \includegraphics[width=\columnwidth]{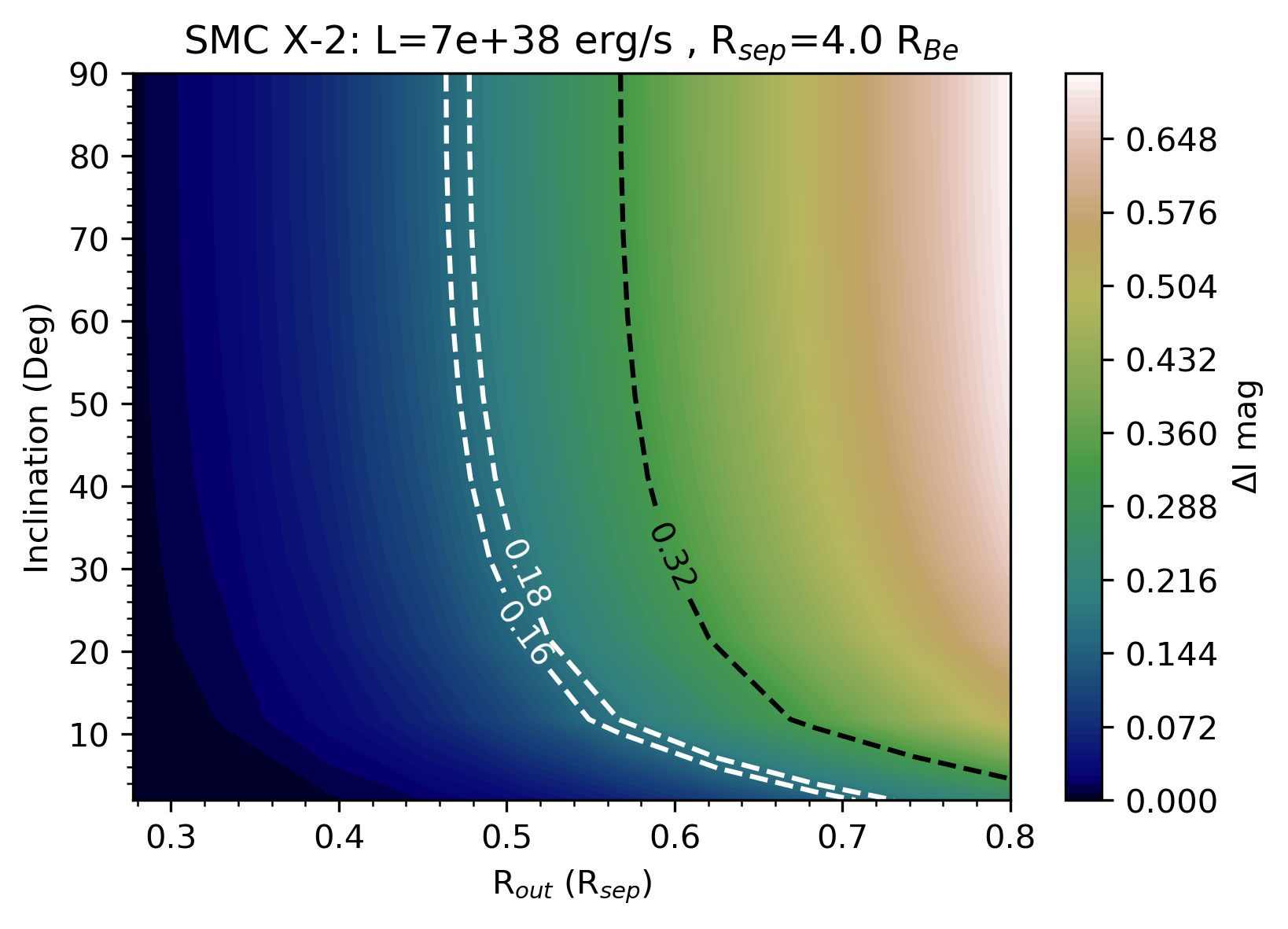}\\
     \includegraphics[width=\columnwidth]{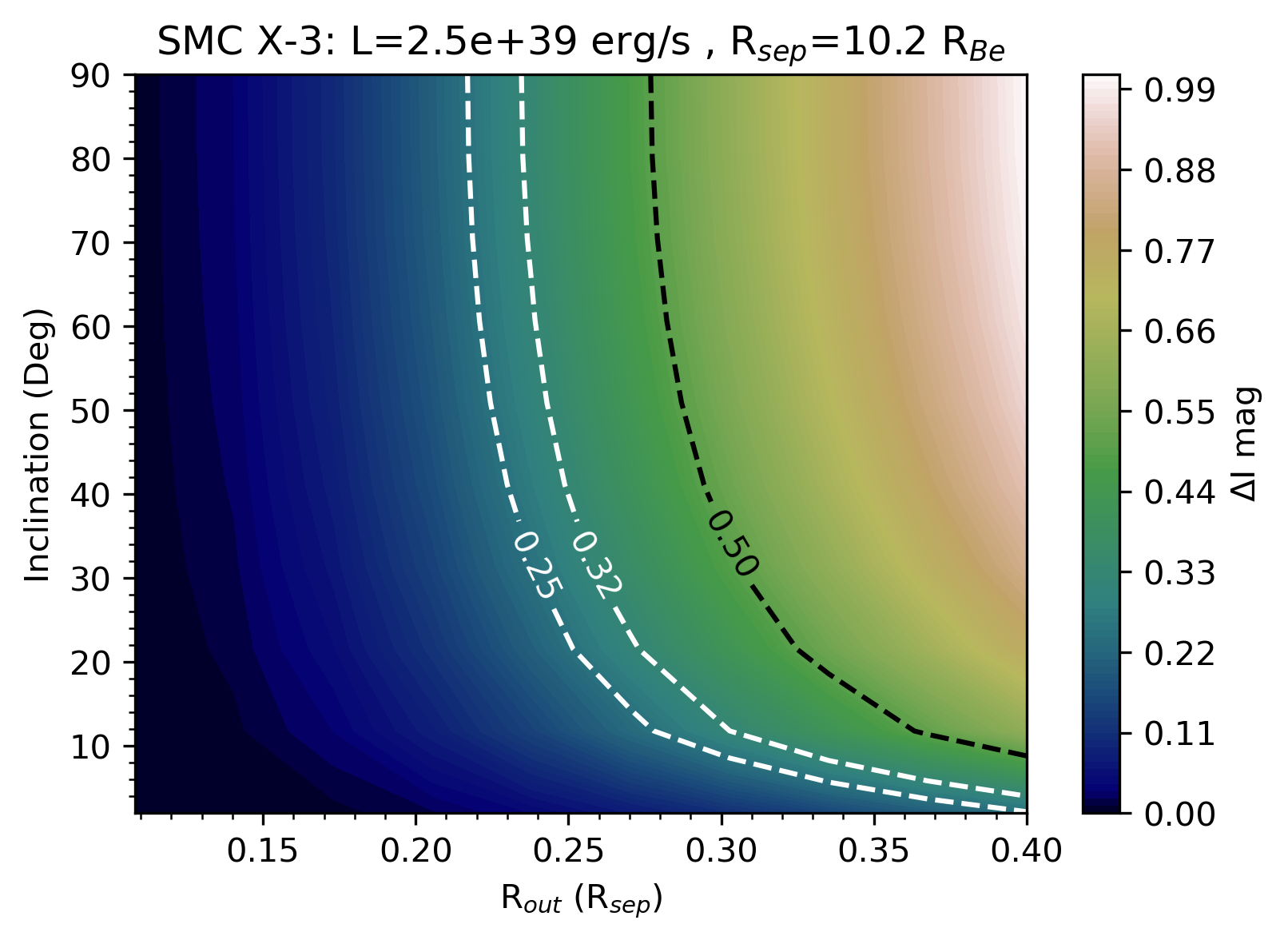}
      \includegraphics[width=\columnwidth]{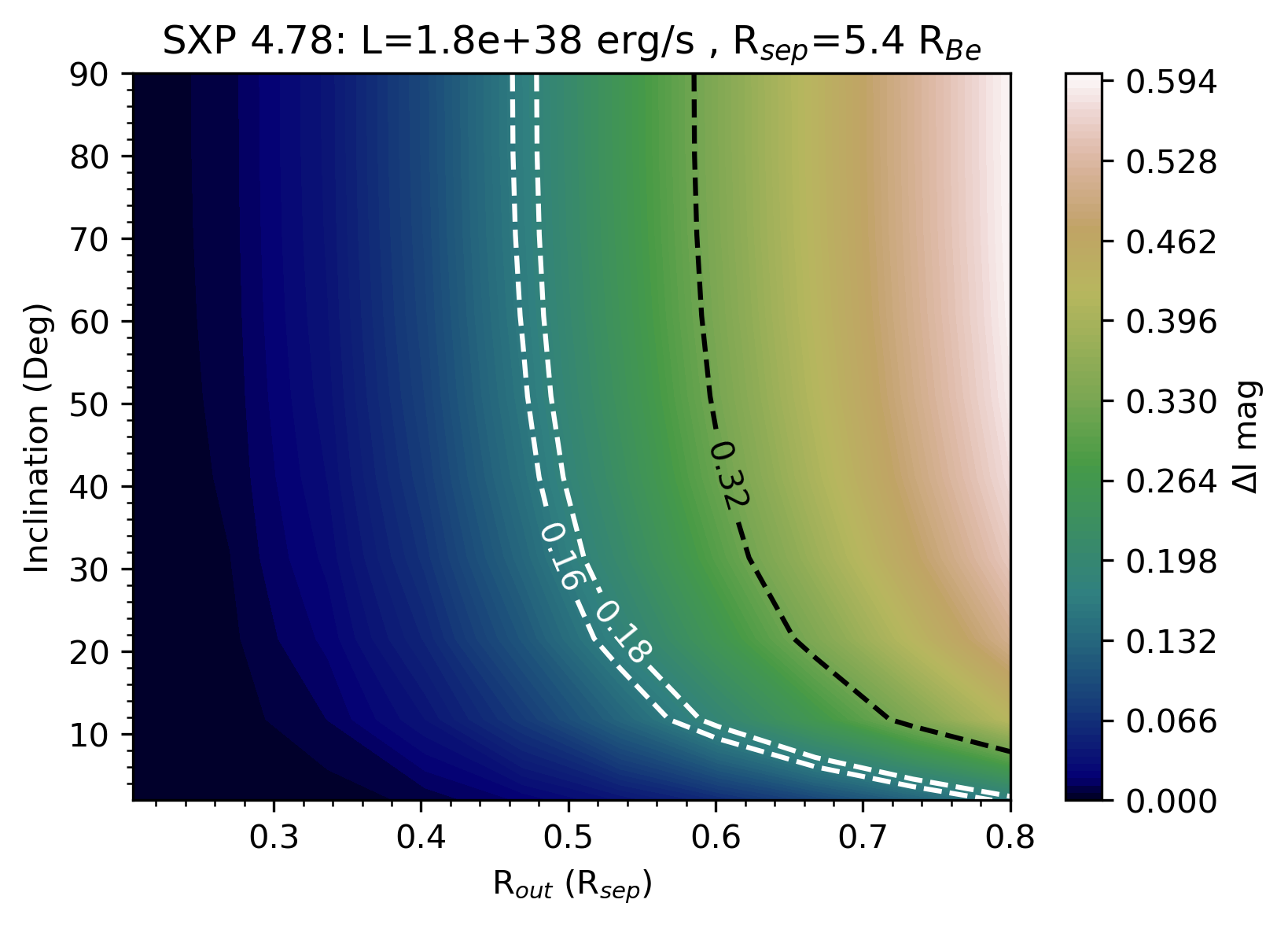}\\
      \includegraphics[width=\columnwidth]{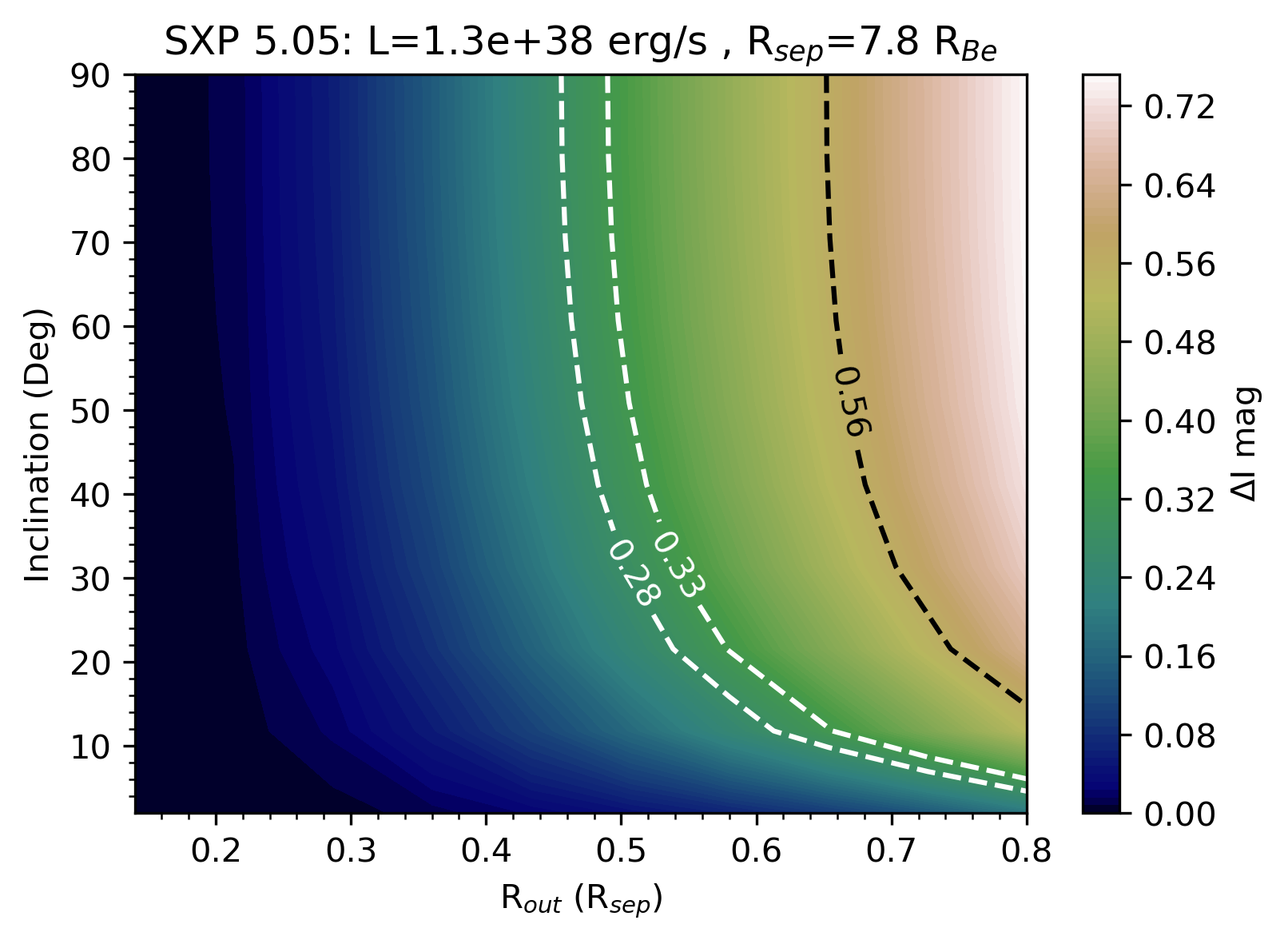}
      \includegraphics[width=\columnwidth]{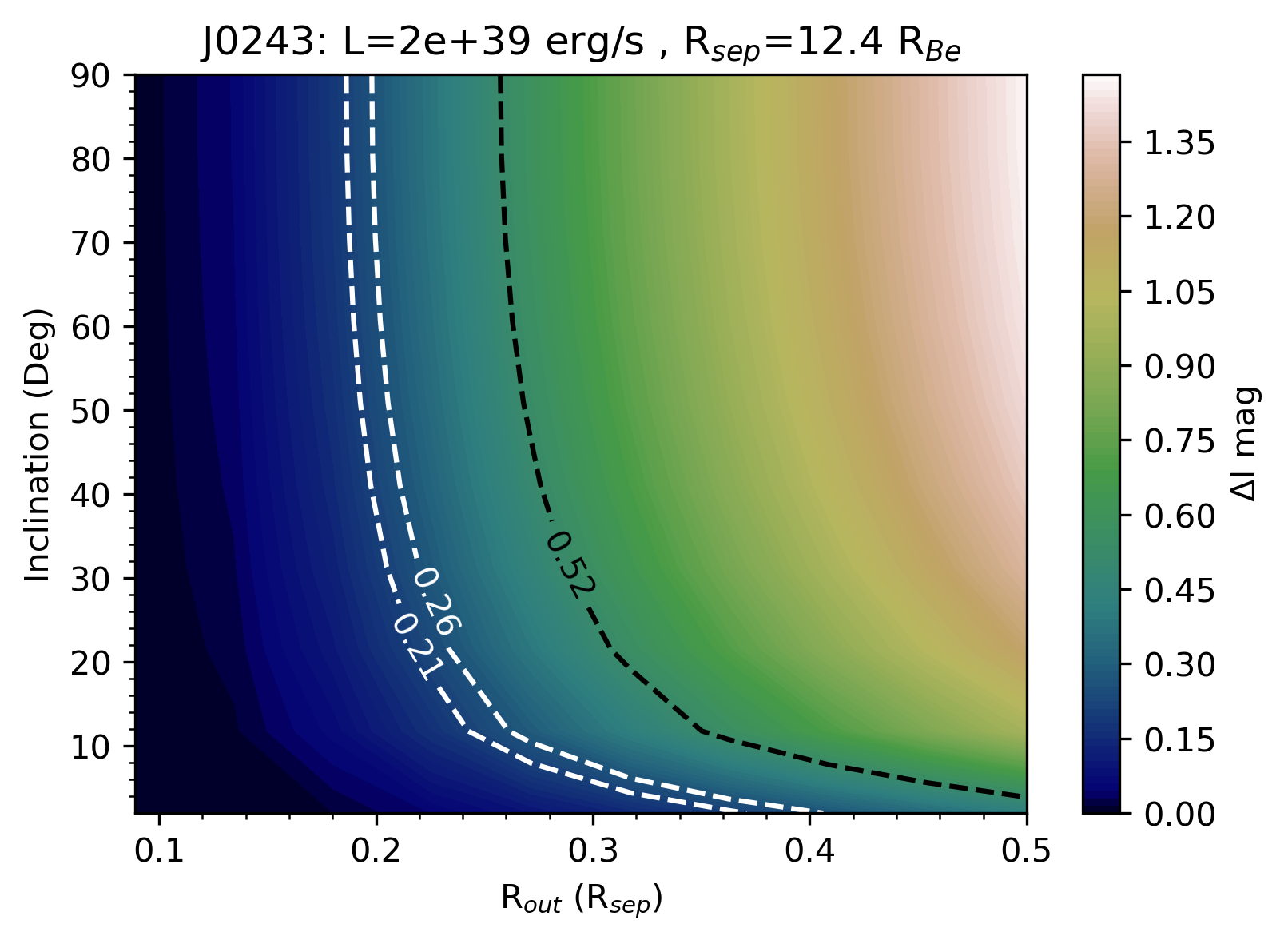}
    \caption{Same as Fig. \ref{fig:heat}, but for a steep temperature profile ($n=2$, $T_{\rm disk,0}=T_\star$ in eq. \ref{eq2}). All the contours are shifted to the left compared to Fig. \ref{fig:heat}.
    }
    \label{fig:heat2}
\end{figure*}

\end{appendix}
\end{document}